\documentclass[lettersize,journal]{IEEEtran}
\usepackage{amsmath,amsfonts}
\usepackage{amsmath,amssymb,amsfonts}
\usepackage{algorithm}
\usepackage{algorithmic}
\usepackage{array}
\usepackage{bm} 
\usepackage[caption=false,font=normalsize,labelfont=sf,textfont=sf]{subfig}
\usepackage{textcomp}
\usepackage{stfloats}
\usepackage{url}
\usepackage{verbatim}
\usepackage{graphicx}
\usepackage{xcolor} 
\usepackage{cite}
\begin{document}

\title{STAR-RIS Assisted Wireless-Powered and Backscattering Mobile Edge Computing Networks}

\author{Bin Lyu,~\IEEEmembership{Member,~IEEE,}~Yining Zhang,~Pengcheng Chen,~\IEEEmembership{Student Member,~IEEE,}~Ziwei Liu, and Feng Tian
}

\markboth{}%
{Shell \MakeLowercase{\textit{et al.}}: A Sample Article Using IEEEtran.cls for IEEE Journals}

\IEEEpubid{}

\maketitle
	\begin{sloppypar}
\begin{abstract}
Wireless powered and backscattering mobile edge computing (WPB-MEC) network is a novel network paradigm to supply energy supplies and computing resource to wireless sensors (WSs). However, its performance is seriously affected by severe attenuations and inappropriate assumptions of infinite computing capability at the hybrid access point (HAP). To address the above issues, in this paper, we propose a simultaneously transmitting and reflecting reconfigurable intelligent surface (STAR-RIS) aided scheme for boosting the performance of WPB-MEC network under the constraint of finite computing capability. Specifically, energy-constrained WSs are able to offload tasks actively or passively from them to the HAP. In this process, the STAR-RIS is utilized to improve the quantity of harvested energy and strengthen the offloading efficiency by adapting its operating protocols. We then maximize the sum computational bits (SCBs) under the finite computing capability constraint. To handle the solving challenges, we first present interesting results in closed-form and then design a block coordinate descent (BCD) based algorithm, ensuring a near-optimal solution. Finally, simulation results are provided to confirm that our proposed scheme can improve the SCBs by 9.9 times compared to the local computing only scheme.
\end{abstract}


\section{Introduction}
\IEEEPARstart{T}he amount of wireless sensors (WSs) and tasks needed to be computed have drastically expanded with the ubiquitous applications of Internet of Things (IoT) technologies~\cite{9751706}. However, to control the manufacturing cost and physical size of WSs, their computing capabilities and battery capacities are generally limited. As a result, they cannot handle complex task requests or sustain a long operating life. Therefore, these WSs cannot efficiently and sustainably support applications with compute-intensive and delay-sensitive requirements, which is a crucial issue limiting the applications of IoT, e.g., virtual reality, autonomous driving, et al~\cite{9698203}. Consequently, it is urgent to provide sufficient computing services and energy supplies for WSs. 
	
Recently, wireless powered mobile edge computing (WP-MEC), a technology that combines wireless power transfer (WPT) with mobile edge computing (MEC), has provided a feasible solution to the above issue~\cite{wang2023wireless}. The fundamental concept underlying WPT is the utilization of energy-carrying radio frequency (RF) signals to supply WSs remotely~\cite{10028982}. Specifically, WSs receive signals from a hybrid access point (HAP) and convert the received signals into electrical energy for supporting their operations. The key idea of MEC is that resource-constrained WSs can offload partial of their tasks to MEC servers deployed nearby~\cite{9940931}. Consequently, WSs can execute tasks by utilizing not only their local computing resources but also the computing resources from the MEC servers. Inspired by the superior advantages above, in a WP-MEC network, WSs can utilize the harvested energy from the HAP for offloading the tasks to the MEC server and implementing the tasks locally by themselves. However, for a typical WP-MEC network, a dedicated phase for implementing the WPT is unavoidable, thereby reducing the time allocated to task offloading and edge computing~\cite{wang2023wireless}. This inherent characteristic seriously limits the system performance and is not conducive to the application of WP-MEC networks. Thus, how to fully explore the utilization of the WPT phase to improve the amount of computed tasks is extremely urgent. 
	
Backscatter Communication (BackCom) has been seen as an attractive option to improve spectral efficiency. For BackCom enabled WSs, they can immediately modulate and reflect the received RF signals for signal delivery~\cite{8327597}. Thus, a dedicated WPT phase is not necessary. Motivated by this, wireless powered and backscattering MEC (WPB-MEC) has been proposed~\cite{9348943}, in which the previous WPT phase can be utilized to support two roles: 1) the passive offloading based on the BackCom module, and 2) the energy harvesting for supporting the future active offloading. Thus, the system spectral efficiency can be improved. However, the signal transmissions between the HAP and multiple WSs may be of poor quality due to serious attenuation, especially for the scenarios where WSs are located in dead zones and cell edges~\cite{9130088}.
	
Recently, reconfigurable intelligent surface (RIS) has attracted increasing attention, which are widely used to assist signal transmissions in MEC systems, aiming to enhance the transmission quality and extend network coverage~\cite{9891794}. An RIS typically consists of three main components: reflection elements, copper boards, and control circuits. Control circuits are used to precisely control the amplitude and phase of each element by adjusting the capacitance and impedance. While copper boards are used for energy conversion and signal transmissions~\cite{di2020smart}. In general, the phase adjustment range of RIS is from 0 to 2$\pi$, which is the basis for realizing an accurate reconstruction of multipath transmissions. Nevertheless, the conventional RIS has a limitation that it is only capable of reflecting incident signals. This implies that in order to achieve effective communications, the transmitter and receiver must be positioned on an identical side of the RIS. This enforces restrictions on the adaptability of utilizing the RIS in different scenarios. In order to get out of this dilemma, simultaneously transmitting and reflecting reconfigurable intelligent surface (STAR-RIS) has received a great of attention~\cite{XuJiaqi2021}. STAR-RIS can reconstruct reflected and transmitted signals for complete spatial coverage by modulating surface impedance. Specifically, STAR-RIS separates the incident signals into two parts: reflected signals and transmitted signals. Reflected signals are reflected into the identical space as the incident signals, i.e., reflection space. While, transmitted signals are transmitted to the opposite space, i.e., transmission space~\cite{9570143}. Due to this excellent performance, it is promising to apply the STAR-RIS in WPB-MEC for enhancing system performance and guaranteeing network feasibility.

\subsection{Related Works}
We undertake a summarization of the available literature from the following perspectives, i.e., WP-MEC networks, WPB-MEC networks, and RIS$\slash$STAR-RIS-assisted MEC networks.

\subsubsection{WP-MEC networks}
The congruent design of WPT and MEC presents an enticing avenue towards self-sustaining ubiquitous computing. \cite{8334188} presented a design to maximize the sum computational bits (SCBs) under the condition that the binary offloading strategy is adopted. In~\cite{9224971}, the authors implemented a collaborative scheme between communication and offloading to minimize the consumed system energy. \cite{8986845} demonstrated the performance gap between the binary and partial offloading modes for WP-MEC networks. In~\cite{9698985}, the authors designed a hybrid offloading strategy for WP-MEC networks. This strategy involves both binary offloading and partial offloading to conserve the residual energy. The authors in~\cite{10210080} explored the full-duplex (FD) technique as a means to achieve the simultaneous energy transfer and task offloading. In \cite{9829184} and \cite{10287987}, the deep reinforcement learning algorithms were designed to maximize the sum computation rate of WP-MEC networks with multiple edge devices.

\subsubsection{WPB-MEC networks}
Compared to WP-MEC networks, the emerging WPB-MEC is more efficient by incorporating the passive offloading and active offloading~\cite{9207959, 9351546, 9812481, 10032541}. 
In~\cite{9207959} and~\cite{9351546}, the computational bits and energy efficiency maximization problems in WPB-MEC networks were investigated, respectively, in which the non-linear energy harvesting (EH) model was utilized. In~\cite{9812481}, the limited computing capability constraint was taken into consideration. \cite{10032541} investigated the integration of WPB-MEC with orthogonal frequency division multiple access.

\subsubsection{RIS$\slash$STAR-RIS-assisted MEC networks}
In recent years, RIS$\slash$STAR-RIS has been widely used in WP-MEC networks as a powerful technology for performance enhancement. In~\cite{9881553}, multiple RISs were placed near the HAP and wireless devices in order to facilitate the WPT and task offloading. In~\cite{9780612}, an RIS aided scheme for WPB-MEC networks was considered, for which the network resource allocation and RIS phase shifts were jointly optimized. In~\cite{Pengcheng}, the impact of time division multiple access (TDMA) and space division multiple access (SDMA) on WP-MEC was investigated. In~\cite{10032506}, the author compared the SCBs with different operating protocols in the STAR-RIS assisted WP-MEC system.

\subsection{Motivations and Contributions}
Recently, extensive efforts have been made to promote the development of WPB-MEC networks. However, there exist some gaps needed to be filled. Firstly, the standard WPB-MEC network generally suffers from the unsatisfied communication efficiency due to severe signal attenuations~\cite{9207959, 9351546, 9812481, 10032541}. Although the reflecting-only RIS was considered as a solution to hackle this problem~\cite{9780612}, it cannot meet the requirement of flexible network deployment since the HAP and WSs have to be placed on the same side of the RIS. Secondly, in the existing works, e.g.,~\cite{9207959, 9351546, 10032541, 9881553, 9780612, 10032506}, the computing capability was assumed to be infinite. This assumption results in the negligence of edge computing time and the mismatch of resource allocation, which is inappropriate for delay-sensitive applications~\cite{9812481}. Thus, there is an urgent need for effective solutions to address the above issues.

In this work, we propose a STAR-RIS aided scheme for the WPB-MEC network with finite computing capability, which comprises an HAP integrated an MEC server, a STAR-RIS and multiple WSs.~The HAP is equipped with multiple antennas for attaining two goals. Firstly, the HAP can support the simultaneous downlink WPT from the HAP to the WSs and uplink passive offloading from the WSs to the HAP by dividing into the antennas into two groups. Secondly, all the antennas are utilized to construct beamforming to receive tasks transferred by active offloading. The STAR-RIS is utilized to guarantee a flexible network deployment, i.e., the HAP and WSs can be deployed on different sides of the STAR-RIS, and assist the downlink WPT and both passive and active offloading for performance improvement.  To exploit the characteristics of the hybrid offloading scheme, the energy splitting (ES) and time splitting (TS) operating protocols are both adopted. Under this setup and  taking the finite computing capability into account, we further investigate the maximization problem of SCBs and obtain an reliable system design.

Compared to~\cite{8334188, 9224971, 8986845, 9698985, 10210080, 9829184, 10287987, 9207959, 9351546, 9812481, 10032541}, the proposed STAR-RIS aided scheme can significantly boost the system performance from the aspects of downlink WPT and uplink hybrid offloading. Unlike~\cite{9881553, 9780612, Pengcheng}, our proposed scheme can guarantee a flexible deployment of WSs and extend the network coverage. Moreover, due to the joint optimization of reflection and transmission coefficients, our formulated problem is much more challenging to solve. In contrast to~\cite{9829184, 10287987, 10032506}, the hybrid offloading scheme can achieve the simultaneous WPT and passive offloading, as well as avoiding interference, thereby improving the utilization of time resource. Also, different from~\cite{10032506}, the consideration of finite computing capability is able to avoid an unreliable system design.

	\begin{itemize}
		\item As far as we know, this is the first time of designing a STAR-RIS aided scheme for WPB-MEC networks. The STAR-RIS is deployed among the WSs for performance improvement. Specifically, it first adopts the ES protocol to assist the downlink WPT and uplink passive offloading. Then, the TS protocol is adopted to aid the uplink active offloading and avoid the interferences. 
		
		\item We formulate an optimization problem of maximizing the SCBs under the finite computing capability constraint. Since the formulated problem is non-convex, we first present closed-form solutions of part of variables by exploiting the network characteristics. Subsequently, we propose a block coordinate descent (BCD) based algorithm, incorporating successive convex approximation (SCA) and semi-positive definite relaxation (SDR) methods, to find the solutions of the remaining variables, which can efficiently address the non-convexity and guarantee the solving accuracy.
		
		\item Simulation results demonstrate the satisfactory convergence performance of our proposed algorithm and confirm that our proposed scheme can enhance the SCBs by 9.9 times compared to the local computing only scheme.
	\end{itemize}

The remaining sections of this paper are structured to come. Section \ref{II} describes the model of a STAR-RIS empowered WPB-MEC network. The formulated problem and proposed solution are shown in Section \ref{III}. Section \ref{IV} presents numerical results for evaluating the proposed algorithm and scheme. Section \ref{V} provides a conclusion.

\section{System Model}\label{II}
\begin{figure}[t]
	\centering\includegraphics[height=6.2cm,width=10cm]{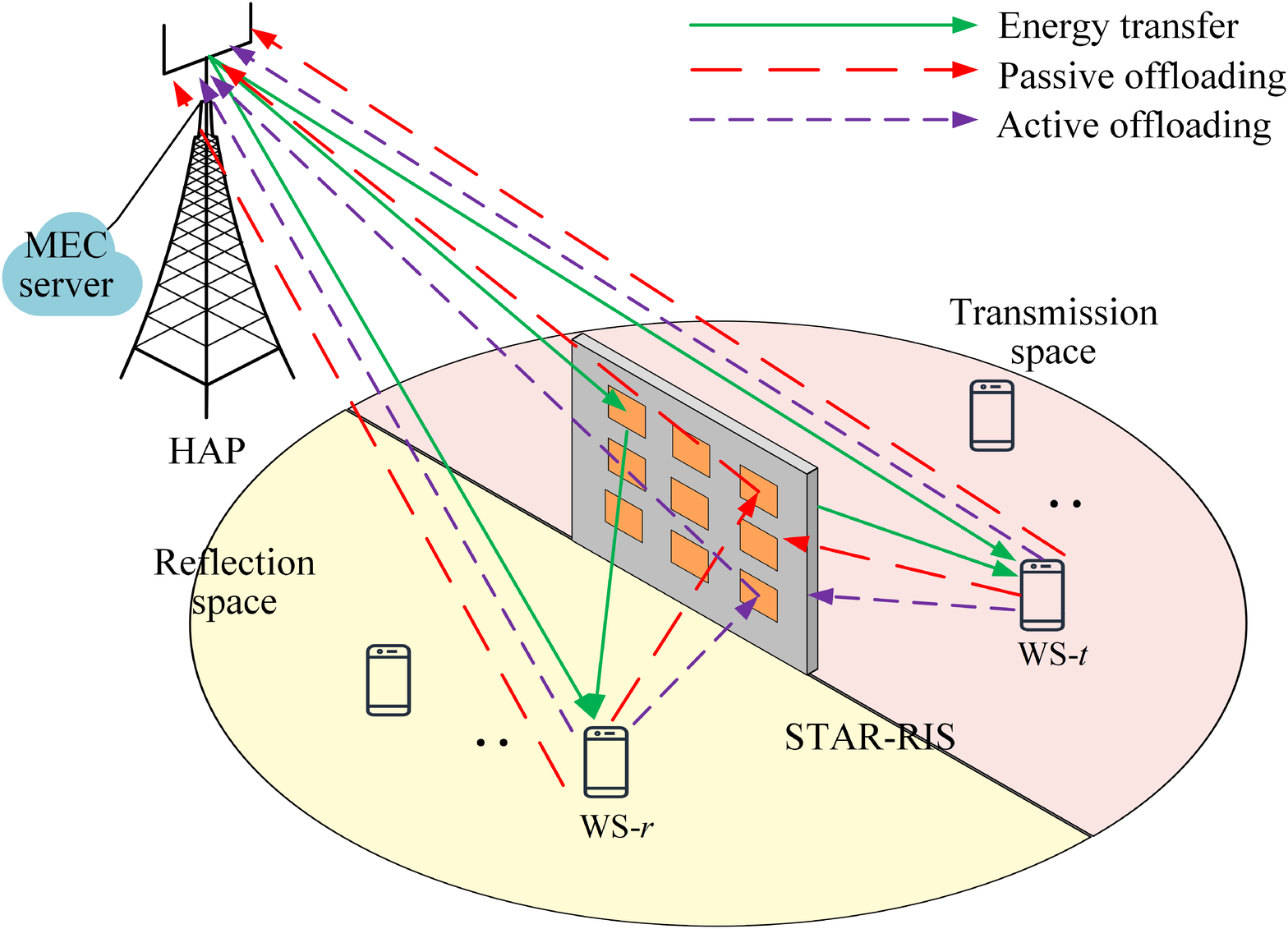}
	\caption{A STAR-RIS enabled WPB-MEC network.}
	\label{system}
\end{figure}

As depicted in Fig.~\ref{system}, we consider a STAR-RIS empowered WPB-MEC network comprising an HAP with $N$ antennas, an MEC server with finite computing capability, a STAR-RIS with $M$ elements, and $K$ single-antenna WSs. The $N$ antennas at the HAP are utilized for supporting the implementation of the FD mode in the downlink and efficient tasks receiving in the uplink\footnotemark[1].\footnotetext[1]{The HAP is also utilized to undertake the control of network scheduling and task computations. As the HAP can be with stable energy supply and high computational capability, the task computations are implemented fast.} Specifically, in the downlink, $\widetilde{N}$ antennas is utilized to transmit energy signals to recharge WSs, and the remaining $\widetilde{N}$ antennas is utilized to simultaneously receive offloaded tasks from WSs on the same frequency band, where $\widetilde{N}=\frac{N}{2}$. However, in the uplink, all the $N$ antennas are explored to construct receive beamforming for receiving the offloaded tasks. Considering that the tasks at the WSs are bit-wise independent and can be arbitrarily partitioned~\cite{8692421}, the WSs employ the partial offloading strategy. Moreover, each WS is equipped with a BackCom module and an active RF component to support both passive offloading and active offloading~\cite{7937935}. Specifically, each WS can offload the tasks to the HAP by either passively backscattering or active transmission. The STAR-RIS is positioned to facilitate the efficient delivery of energy from the HAP to all WSs in the downlink and passive and active offloading from the WSs to the HAP in the uplink. According to the location of the STAR-RIS, the communication space is divided into reflection space and transmission space. Without loss of generality, we consider that there are $R$ WSs and $T$ WSs in the reflection and transmission spaces, respectively, where $R+T=K$. Accordingly, the sets of the WSs in the reflection and transmission spaces are denoted by $\mathcal{R}=\{1,\ldots, R\}$ and $\mathcal{T}=\{1, \ldots, T\}$, respectively.

\begin{figure}[t]
	\centering\includegraphics[height=5.5cm,width=8cm]{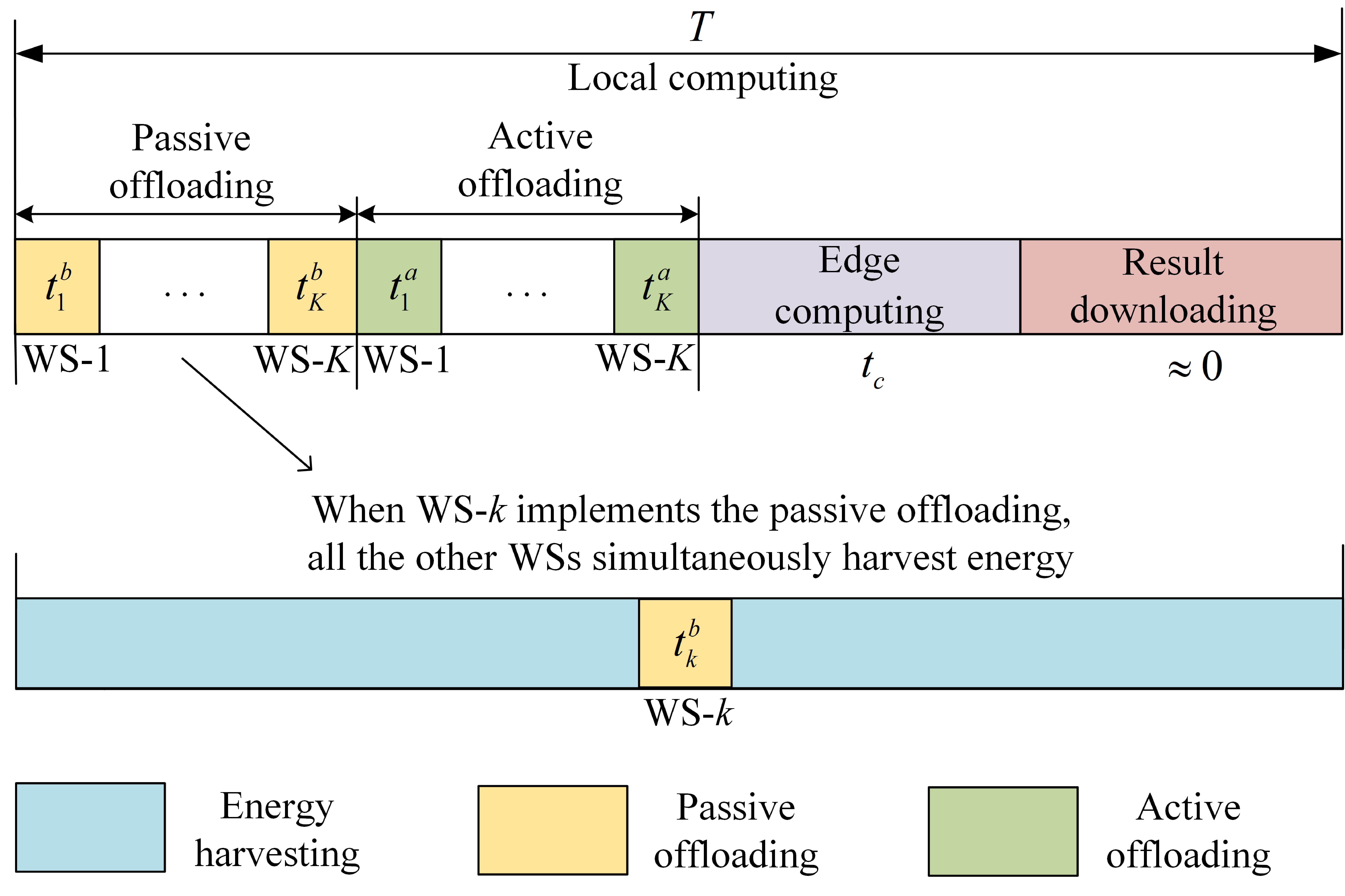}
	\caption{The illustration of time allocation.}
	\label{time}
\end{figure}

As depicted in Fig. \ref{time}, the transmission block under consideration has a duration of $T$ seconds and consists of four phases, i.e., passive offloading phase, active offloading phase, edge computing phase and result downloading phase.~Nevertheless, considering the relatively small size of computation results, it is reasonable to neglect the duration required for downloading results from the HAP (i.e. MEC server) to WSs~\cite{10210080}. Thus, we focus on the operations of the first three phases in the following. Moreover, thanks to advanced power distribution techniques~\cite{power}, each WS can compute its own tasks locally throughout the block~\cite{9207959, 9881553}. This arises from the capability of utilizing the harvested energy to power the chip processors at each WS~\cite{9881553}, which will be described in Section \ref{II-D}.

\subsection{Passive Offloading Phase}
In the passive offloading phase, the HAP leverages the FD mode to transmit energy signals to the WSs and receive the tasks offloaded by the WSs at the same time. To enable the FD operation, the ES operating protocol of STAR-RIS is adopted in this phase. The amplitudes for reflection and transmission of the $m$-th element of the STAR-RIS are represented by $\beta_{m}^r$ and $\beta_{m}^t$, respectively, where $\beta_{m}^r$ + $\beta_{m}^t$ = 1 and $\beta_{m}^r$, $\beta_{m}^t\in[0,1]$. The phase shifts for reflection and transmission of the $m$-th element are represented by $\theta_{m}^r$ and $\theta_{m}^t$, respectively, where $\theta_{m}^r$, $\theta_{m}^t\in[0,2\pi)$\footnotemark[2].\footnotetext[2]{Similar to \cite{9570143, 10032506, 10153701}, we consider the phase shifts for reflection and transmission can be adjusted independently by using the equivalent surface electric and magnetic currents.} Therefore, the coefficient matrices of the STAR-RIS in this phase are given by $\mathbf{\Theta}_{x}=\mathrm{diag}(\sqrt{\beta_{1}^x}e^{j\theta_{1}^x}, \sqrt{\beta_{2}^x}e^{j\theta_{2}^x}, \cdots, \sqrt{\beta_{M}^x}e^{j\theta_{M}^x}),~x\in\{r,t\}$, $r\in\mathcal{R}$, and $t\in\mathcal{T}$. To avoid interference for passive offloading, we adopt the TDMA protocol, which divides this phase into $K$ sub-phases. During the $k$-th sub-phase with $t_{k}^b$ seconds, the WS-$k$ passively backscatters its task to the HAP. In the meanwhile, the other WSs simultaneously harvest energy from the HAP.

The transmit signal at the HAP during $t_{k}^b$ is given by
	\begin{equation}
		\bm{x}_{\textit{p}}=s(t)\bm{\omega}_{0,k}, 
	\end{equation}
	where $s(t)$ is the information-carrying signal with unit power, and $\bm{\omega}_{0,k}\in\mathbb{C}^{\widetilde{N}\times 1}$ denotes the transmit beamforming vector of the HAP, which satisfies $\sum\limits_{k=1}^{K}\left\|{\bm{\omega}}_{0,k}\right\|^2\le {{P}_{t}}$, and ${{P}_{t}}$ denotes the maximum transmit power at the HAP.~The received signal at the WS-$k$ is expressed as
	\begin{equation}
		\begin{aligned}
			y_{{k}}&=\left(\bm{g}_{\textit{s,x}}^H\bm{\Theta}_{x}\bm{G}_{\textit{p,s}}+\bm{g}_{\textit{p,x}}^H\right)\bm{x}_{\textit{p}}+{n}_{k}\\
			&=\bm{g}_k^Hs(t)\bm{\omega}_{0,k}+{n}_{k},
		\end{aligned}
	\end{equation}
	where $\bm{g}_k^H\triangleq\bm{g}_{\textit{s,x}}^H\bm{\Theta}_{x}\bm{G}_{\textit{p,s}}+\bm{g}_{\textit{p,x}}^H$ represents the downlink channel coefficient\footnotemark[3],\footnotetext[3]{In this paper, we assume all CSI can be perfectly estimated and focus on the study of the upper-bound of system performance. In the future work, we will extend to study how the imperfect CSI affects system performance.} $\bm{g}_{\textit{p,x}}^H\in\mathbb{C}^{1 \times\widetilde{N} }$ , $\bm{G}_{\textit{p,s}}\in\mathbb{C}^{M\times\widetilde{N}}$and $\bm{g}_{\textit{s,x}}^H\in\mathbb{C}^{1\times M}$ represent the downlink channels from the HAP to the WS-$k$, from the HAP to the STAR-IRS, and from the STAR-RIS to the WS-$k$, respectively, and ${n}_{k}\sim\mathcal{CN}(0,\sigma_k^2)$ represents the additive white Gaussian noise (AWGN) at the WS-$k$. The reflected signal at the WS-$k$ is represented by
	\begin{equation}\label{x_k}
		\begin{aligned}
			x_{{k}}&=\sqrt{\rho_k}e_k(t)y_{{k}}+n_{k}\\
			&=\sqrt{\rho_k}e_k(t)\bm{g}_k^Hs(t)\bm{\omega}_{0,k}+n_{k}.
		\end{aligned}
	\end{equation}
	In (\ref{x_k}), $e_k(t)$ with unit power indicates the offloading symbol of the WS-$k$, and $\rho_k$ represents the power reflection coefficient of the WS-$k$ with $0\leq \rho_k\leq 1$. According to~\cite{9417430}, all reflection coefficients are assumed to identical due to the setting of hardware controllers, i.e, $\rho_k=\rho$. Thus, during $t_k^b$, the received signal at the HAP is
	\begin{equation}
		\begin{aligned}
			y_{\textit{c}}^b=&\sqrt{\rho}e_k(t)\bm{\omega}_{1,k}^H\bm{h}_k\bm{g}_k^Hs(t)\bm{\omega}_{0,k}\\
			&+\bm{\omega}_{1,k}^H(\bm{H}_{\textit{s,c}}^H\bm{\Theta}_{x}\bm{G}_{\textit{p,s}}+\bm{b})s(t)\bm{\omega}_{0,k}+\bm{\omega}_{1,k}^H\bm{n}_{\textit{c}},
		\end{aligned}
	\end{equation}
	where $\bm{h}_k\triangleq\bm{H}_{\textit{s,c}}^H\bm{\Theta}_{x}\bm{h}_{\textit{s,x}}+\bm{h}_{\textit{c,x}}$ represents the uplink channel coefficient from the WS-$k$ to the HAP, $\bm{h}_{\textit{c,x}}\in\mathbb{C}^{\widetilde{N}\times 1}$, $\bm{H}_{\textit{s,c}}^H\in\mathbb{C}^{\widetilde{N}\times M}$ and $\bm{h}_{\textit{s,x}}\in\mathbb{C}^{M\times 1}$ represent the uplink channels from the WS-$k$ to the HAP, from the STAR-IRS to the HAP, and from the the WS-$k$ to the STAR-RIS in the passive offloading phase, respectively, and $\bm{\omega}_{1,k}^H\in\mathbb{C}^{1\times \widetilde{N}}$ is the receive beamforming vector at the HAP in the passive offloading phase. In (4), the first term $\sqrt{\rho}e_k(t)\bm{\omega}_{1,k}^H\bm{h}_k\bm{g}_k^Hs(t)\bm{\omega}_{0,k}$ is the desired signal, the second term $\bm{\omega}_{1,k}^H(\bm{H}_{\textit{s,c}}^H\bm{\Theta}_{x}\bm{G}_{\textit{p,s}}+\bm{b})s(t)\bm{\omega}_{0,k}$ is the self-interference (SI), and $\bm{n}_{\textit{c}}\sim\mathcal{CN}(0,\sigma^2\bm{I}_{\widetilde{N}})$ is the AWGN at the HAP. It is observed that the SI consists of two parts. To be specific, $\bm{\omega}_{1,k}^H(\bm{H}_{\textit{s,c}}^H\bm{\Theta}_{x}\bm{G}_{\textit{p,s}})s(t)\bm{\omega}_{0,k}$ is the interference caused by the reflected link, and $\bm{\omega}_{1,k}^H\bm{b}s(t)\bm{\omega}_{0,k}$ is the interference caused by the HAP's loopback link. However, $\bm{\omega}_{1,k}^H(\bm{H}_{\textit{s,c}}^H\bm{\Theta}_{x}\bm{G}_{\textit{p,s}})s(t)\bm{\omega}_{0,k}$ can be suppressed after the coefficient optimization at the STAR-RIS \cite{SI}. Thus, according to \cite{10210080}, the second term of (4) can be approximated as 
	\begin{equation}\label{SI}
		\bm{\omega}_{1,k}^H(\bm{H}_{\textit{s,c}}^H\bm{\Theta}_{x}\bm{G}_{\textit{p,s}}+\bm{b})s(t)\bm{\omega}_{0,k}\approx\bm{\omega}_{1,k}^H\bm{b}s(t)\bm{\omega}_{0,k}.
	\end{equation}
	
	Since $s(t)$ is previously known at the HAP, the SI expressed in (\ref{SI}) is further diminished by applying self-interference cancellation (SIC) techniques in both analog and digital domains. After that, the remaining SI power is characterized by $\sqrt{\beta\gamma}\bm{\omega}_{1,k}^H\bm{b}s(t)\bm{\omega}_{0,k}$, where $\beta$ and $\gamma$ denote the scaling factors by applying SIC techniques. Therefore, the computational bits offloaded by the WS-$k$ through passively backscattering are expressed as
	\begin{equation}\label{Rkb}
		R_{k}^b=t_k^bB\log_2\left(1+\dfrac{\xi\rho|\bm{\omega}_{1,k}^H\bm{h}_k\bm{g}_k^H\bm{\omega}_{0,k}|^2}{\beta\gamma|\bm{\omega}_{1,k}^H\bm{b}\bm{\omega}_{0,k}|^{2}+\sigma^2\left\|\bm{\omega}_{1,k}\right\|^2}\right), 
	\end{equation}
	where $B$ is the channel bandwidth, and $\xi$ is the performance gap caused by the practical modulation used in the passive offloading phase~\cite{7981380}. Note that the power consumption of WSs for passive offloading is not taken into account in (\ref{Rkb}). This is due to the fact that the power consumed by passively backscattering is much less than the received power and is thus negligible~\cite{7937935}.
	
	During $t_{k}^b$, a part of the received power at the WS-$k$, i.e., $\rho|{\bm{g}_k^H}\bm{\omega}_{0,k}|^2$, is consumed to enable the passive offloading, while the remainder, i.e., $\left(1-\rho\right)|{\bm{g}_k^H}\bm{\omega}_{0,k}|^2$, enters the EH circuit for harvesting energy. In the meanwhile, the WS-$i$ $\left({i}\in\mathcal{K}\backslash{k}\right)$ focuses on harvesting energy from the HAP. According to the non-linear EH model \cite{7264986}, the amount of energy harvested by the WS-$k$ during the passive offloading phase is formulated as
	\begin{equation}
		E_k=E_k^b+\sum_{i=1,i\neq{k}}^{K}t_i^bP_{k,i}^b,
	\end{equation}  
	where $E_{k}^b\!\!=\!\! t_k^b\!\left(\!\!\dfrac{X_k}{1+\exp\left(-a_k\left(\left(1-\rho \right)|{\bm{g}_k^H}\bm{\omega}_{0,k}|^2-b_k \right) \right)}\right.\\
	\left.-Y_k\right)$, $P_{k,i}^b\!\!=\!\!\left(\dfrac{X_k}{1+\exp\left(-a_k\left(|{\bm{g}_k^H}\bm{\omega}_{0,i}|^2-b_k \right) \right)}-Y_k\right)$, $X_k=\frac{{P_{\rm{sat}}\left(1+\exp\left({a_{k}b_{k}}\right)\right)}}{{\exp\left({a_{k}b_k}\right)}}$, $Y_k=\frac{{P_{\rm{sat}}}}{{\exp\left({a_{k}b_k}\right)}}$. $P_{\rm{sat}}$ is the saturation power, $a_k$ and $b_k$ are the parameters controlled by the EH circuit's construction at the WS-$k$~\cite{7264986}.

\subsection{Active Offloading Phase}
In the active offloading phase, the HAP stops transmitting energy signals, and $K$ WSs sequentially offload their tasks to the HAP with the harvested energy in the passive offloading phase. During this phase, the HAP's $N$ antennas are both used for receiving the offloaded tasks. To facilitate the utilization of the TDMA protocol, i.e., the active offloading phase is divided into $K$ sub-phases, and the TS operating protocol is adopted for the STAR-RIS. The reflection and transmission amplitudes of the $m$-th element of the STAR-RIS for the $k$-th sub-phases are represented by $\beta_{m}^{r,k}$ and $\beta_{m}^{t,k}$, respectively. According to the characteristics of the TS protocol, all elements work either in the transmission mode or in the reflection mode, i.e., $\beta _{m}^{t,k}=1$ and $\beta _{m}^{r,k}=0$ if the transmission mode is activated, or $\beta _{m}^{t,k}=0$ and $\beta _{m}^{r,k}=1$ if the reflection mode is activated. The phase shifts for reflection and transmission of the $m$-th element are represented by $\theta_{m}^{r,k}$ and $\theta_{m}^{t,k}$, respectively, where $\theta_{m}^{r,k}$, $\theta_{m}^{t,k}\in[0,2\pi)$. Thus, the coefficient matrices of the STAR-RIS in this phase are expressed as $\mathbf{\Theta}_{x,k}=\mathrm{diag}(\sqrt{\beta_{1}^{x,k}}e^{j\theta_{1}^{x,k}}, \sqrt{\beta_{2}^{x,k}}e^{j\theta_{2}^{x,k}}, \cdots, \sqrt{\beta_{M}^{x,k}}e^{j\theta_{M}^{x,k}})$, $x\in\{r,t\}$. During the $k$-th sub-phase with $t_k^a$ seconds, the WS-$k$ actively transmits its tasks to the HAP. Let $p_k$ represent the transmit power of the WS-$k$, and then the signal received by the HAP from the WS-$k$ during $t_k^a$ is given by

\begin{equation}
	\begin{aligned}\label{8}
		y_{\textit{c}}^a&=\bm{\omega}_{2,k}^{H}\left(\bm{H}_{{s,c,2}}^H\bm{\Theta}_{x,k}\bm{h}_{\textit{s,x}}+\bm{h}_{\textit{c,x,2}}\right)\sqrt{p_k}e_k(t)\!\!+\!\!\bm{\omega}_{2,k}^{H}\bm{n}_{\textit{c}}\\
		&=\bm{\omega}_{2,k}^H\bm{h}_{k,2}\sqrt{p_k}e_k(t)+\bm{\omega}_{2,k}^{H}\bm{n}_{\textit{c}},
	\end{aligned}
\end{equation}where $\bm{h}_{k,2}\triangleq\bm{H}_{\textit{s,c,2}}^H\bm{\Theta}_{x,k}\bm{h}_{\textit{s,x}}+\bm{h}_{\textit{c,x,2}}$ represents the uplink channel coefficient from the WS-$k$ to the HAP in the active offloading phase, $\bm{h}_{\textit{c,x,2}}\in\mathbb{C}^{N\times 1}$ and $\bm{H}_{\textit{s,c,2}}^H\in\mathbb{C}^{N\times M}$ denote the uplink channels from the WS-$k$ to the HAP and from the STAR-IRS to the HAP, respectively, and $\bm{\omega}_{2,k}^H\in\mathbb{C}^{1\times N}$ is the receive beamforming vector with unit power at the HAP in the active offloading phase. The calculation of the computational bits offloaded by the WS-$k$ during $t_k^a$ is given by 
\begin{equation}\label{Rka}
	R_{k}^a=t_k^aB\log_2\left(1+\dfrac{p_k|\bm{\omega}_{2,k}^H\bm{h}_{k,2}|^2}{\sigma^2\left\|\bm{\omega}_{2,k}\right\|^2 } \right). 
\end{equation} 

In this phase, the energy consumption for enabling the active offloading of the WS-$k$ is $ {{p}_{k}}{{t}_{k}^a}+E_{c,k}$, where $E_{c,k}$ is the constant circuit energy consumption.

\subsection{Edge Computing Phase}
In the edge computing phase, the HAP and all WSs remain silent, while the MEC server initiates tasks computing upon their reception in the passive and active offloading phases. Let $f_m$ (in Hz) and $t_c$ (in seconds) represent the computing frequency and the duration of edge computing phase for the MEC server, respectively. To ensure that the MEC server is capable of fully processing all received tasks, the subsequent constraint should be satisfied, i.e.,
\begin{equation}\label{10}
	\sum\limits_{k=1}^{K}{(R_{k}^{b}+R_{k}^{a}){\varphi_k}}\le {{f}_{m}}{{t}_{c}},
\end{equation}
where ${\varphi_k}$ represents the number of cycles consumed for computing one bit. 


	\subsection{Local Computing}\label{II-D}
	Let $f_k$ (in Hz) and $\tau_k$ (in seconds) represent the computing frequency and time of the WS-$k$, respectively. The computational bits for the WS-$k$'s local computing are given by
	\begin{equation}\label{Rek}
		R_{k}^{e}=\frac{{{\tau}_{k}}{{f}_{k}}}{{{\varphi }_{k}}}.
	\end{equation}
	
	Local computing energy consumption at the WS-$k$ can be calculated by
	\begin{equation}
		E_{k}^{e}={{\varepsilon }_{k}}f_{k}^{3}{{\tau}_{k}},
	\end{equation}
	where ${\varepsilon }_{k}$ is the effective capacitance coefficient of the on-chip processor at the WS-$k$.
	
\section{Computational Bits Maximization} \label{III}
In this section, we investigate the maximization problem of the SCBs by jointly optimizing the network time allocation, the transmit and receive beamforming designs of the HAP, the computing frequency of the MEC server, the amplitudes and phase shifts of the STAR-RIS for both passive and active offloading phases, the computing time, the computing frequencies and the transmit power of WSs. Let $\bm{t}=\{{t}_k^b, {t}_k^a, t_c, k\in\mathcal{K}\}$, $\bm{\omega}=\{\bm{\omega}_{0,k},\bm{\omega}_{1,k},\bm{\omega}_{2,k}, k\in\mathcal{K}\}$, $\bm{\tau}=\{\tau_k, k\in\mathcal{K}\}$, $\bm{f}=\{f_k, k\in\mathcal{K}\}$, and $\bm{p}=\{p_{k}, k\in\mathcal{K}\}$. Following the aforementioned descriptions, the formulation of the SCBs maximization problem is given by
\begin{align*}  
	(\mathbf{P0}):&\max_{\bm{u}}~{\sum\limits_{k=1}^{K}{(R_{k}^{e}+R_{k}^{b}+R_{k}^{a})}} \\ 
	\mathrm{s.t.~}&\mathrm{C1}:~{p}_{k}t_{k}^{a}+E_{c,k}+{{\varepsilon }_{k}}f_{k}^{3}{{\tau}_{k}}\le {{E}_{k}},~\forall k,\\ 
	&\mathrm{C2}:~\sum\limits_{k=1}^{K}\left\|{\bm{\omega}}_{0,k}\right\|^2\le {{P}_{t}},~\forall k,\\ 
	&\mathrm{C3}:~0\le {{f}_{m}}\le {{f}_{\max}}, \\
	&\mathrm{C4}:~0\le {{f}_{k}}\le {{f}_{k}^{\max}}, ~\forall k,\\ 
	&\mathrm{C5}:~\sum\limits_{k=1}^{K}{\left( t_{k}^{b}\!+\! t_{k}^{a} \right)\!+\!{{t}_{c}}\!\le\! T}, t_{k}^{b},t_{k}^{a},{{t}_{c}}\ge 0, ~\forall k,\\ 
	& \mathrm{C6}:~0\le{{\tau}_{k}}\le{T},~ \forall k,\\ 
	&\mathrm{C7}:~0\le{p}_k,~ \forall k,\\ 
	& \mathrm{C8}:~\beta _{m}^{t}+\beta _{m}^{r}=1,~ \forall m\in \mathcal{M}, \\ 
	&\qquad\ {0}\le \beta _{m}^{t},\beta _{m}^{r}\le 1, ~\forall m\in \mathcal{M}, \\  	
	&\mathrm{C9}:~\sum\limits_{k=1}^{K}{(R_{k}^{b}+R_{k}^{a}){\varphi_k}}\le {{f}_{m}}{{t}_{c}},~ \forall k,
\end{align*}
where $\bm{u}=\{\bm{t}, \bm{\omega},  f_m, \mathbf\Theta_x, \{\mathbf\Theta_{x,k}\}_{k=1}^K , \bm{\tau}, \bm{f}, \bm{p}\}$, ${f}_{\max}$ and ${f}_{k}^{\max}$ are the maximum computing frequencies for the MEC server and WS-$k$, respectively.

In ($\mathbf{P0}$), $\mathrm{C1}$ is the energy causality constraint guaranteeing that the energy consumed by each WS does not exceed its harvested energy. $\mathrm{C2}$ is the transmit power constraint at the HAP. $\mathrm{C5}$ and $\mathrm{C9}$ constrain the network time allocation, while $\mathrm{C8}$ is the energy conservation constraint on each STAR-RIS element. $\mathrm{C3}$, $\mathrm{C4}$, $\mathrm{C6}$, and $\mathrm{C7}$ are the non-negative constraints about ${f}_m$, $\bm{f}$, $\bm{\tau}$, and $\bm{p}$, respectively.

The coupling of variables in both the objective function and the constraints result in the non-convexity of ($\mathbf{P0}$), which thus cannot be solved by convex optimization techniques. 

Before solving ($\mathbf{P0}$), we first analyze the structure of ($\mathbf{P0}$) and derive the following interesting results: 
\begin{itemize}
	\item $f_m^*=f_{\max}$, $\tau_k^*=T$. We can clearly observe from (\ref{10}) that when the MEC server operates at a lower computing frequency, it takes more computing time to process tasks. Thus, the MEC server prefers to compute tasks with its maximum frequency, thereby saving more time allocated to passive and active offloading. Meanwhile, we can also observe that $R_k^e$ formulated in (\ref{Rek}) is a monotonically increasing function of $\tau_k$. From this, we conclude that the maximum SCBs are attained when the MEC server performs all received tasks at its maximum computing frequency, and each WS performs local computing within the entire block. 
		\item \begin{equation}\label{w1}
			{\bm{\omega}_{1,k}}^{*}=\mathfrak{A}^{-1}\mathfrak{B},
		\end{equation} where $\mathfrak{A}=\bm{h}_k\bm{g}_k^H\bm{\omega}_{0,k}\in\mathbb{C}^{\widetilde{N}\times 1}$, $\mathfrak{B}=\beta\gamma\bm{b}\bm{\omega}_{0,k}\bm{\omega}_{0,k}^H\bm{b}^H+\sigma^2\bm{I}_{\widetilde{N}}$. Note that the optimization variable $\bm{\omega}_{1,k}$ only appears in (\ref{Rkb}). For given $\bm{\omega}_{k}$ and $\mathbf\Theta_x$, the closed solution of $\bm{\omega}_{1,k}$ can be given as (\ref{w1})~\cite{5447076}.
	\item \begin{equation}\label{w}
		{\bm{\omega}_{2,k}}^{*}=\frac{\sqrt{p_k}{{\bm{h}}_{k,2}}}{\sigma^2}.
	\end{equation} As the TDMA protocol is utilized in the active offloading phase, the design of receive beamforming of each sub-phase is independent of other variables. The optimal solution, denoted by ${\bm{\omega}_{2,k}}^{*}$, is attained by exploring the maximum ratio combing (MRC) \cite{Pengcheng}. Compared to the scenario with a single-antenna receiver \cite{9780612} or without designing receive beamforming, the result shown in (\ref{w}) can efficiently strengthen the received signal power, guaranteeing that more tasks can be offloaded to the HAP for edge computing.
	\item \begin{equation}\label{c}
		\begin{aligned}	
			{(\bm{\theta}^{x,k})}^*&=\arg \left( {{\bm{\omega}}_{{2,k}}^H}\bm{h}_{\textit{c,x,2}} \right) \\
			&-\arg \left( \mathrm{diag}\left({{\bm{\omega}}_{{2,k}}^H}\bm{H}_{\textit{s,c,2}}^H\right) \bm{h}_{\textit{s,x}}\right),
		\end{aligned}
	\end{equation}where $\bm{\theta}^{x,k}=\{{\theta}_1^{x,k},\cdots,{\theta}_m^{x,k},\cdots,{\theta}_M^{x,k}\}$, and arg$(x)$ is the phase of $x$. Similar to the analysis of receive beamforming, the design of STAR-RIS phase shifts in the active offloading phase is also independent of other variables. Thus, we can derive the optimal phase shifts in the active offloading phase as shown in (\ref{c}) \cite{9214497}. It is notable that the setting of reflection and transmission amplitudes in this phase is controlled by the channel conditions.
\end{itemize}

With the above results, we substitute $f_m=f_m^*$, $\tau_k^*=T$, ${\bm{\omega}_{1,k}}={\bm{\omega}_{1,k}}^{*}$, ${\bm{\omega}_{2,k}}={\bm{\omega}_{2,k}}^{*}$ and $\mathbf\Theta_{x,k}=\mathbf\Theta_{x,k}^{*}$ into ($\mathbf{P0}$). However, these results cannot address the non-convexity completely and other techniques need to be further implemented. In order to address the coupling term of ${p}_{k}$ and $t_{k}^{a}$ in $\mathrm{C1}$, an auxiliary variable, defined as ${q}_{k}=t_{k}^{a}{p}_{k}$, is introduced, based on which ${p}_{k}$ can be equivalently represented by $\frac{{q}_{k}}{t_{k}^{a}}$. To simplify the notations, we then let ${{ \varpi }_{k}^{*}}\triangleq\frac{{\left|\bm{\omega}_{2,k}^H\bm{h}_{k,2} \right|}^{2}}{\left\|\bm{\omega}_{2,k}\right\|^2}$  and ${R_{\rm{sum}}}\triangleq\sum\limits_{k=1}^{K}\frac{T{f}_{k}}{{\varphi}_{k}}+\sum\limits_{k=1}^{K}\left(R_{k}^{b}+Bt_{k}^{a}{{\log}_{2}}\!\left(1\!+\frac{{{q}_{k}}{{ \varpi }_{k}^{*}}}{{{{t}_{k}^a\sigma }^{2}}}\!\right)\right)$. Then, ($\mathbf{P0}$) is simplified as
\begin{align*} 	
	(\mathbf{P1}):&\max_{\bm{t}, \mathbf\Theta_x ,\bm{f},\bm{p},\{\bm{\omega}_{0,k}\}_{k=1}^K}{R_{\rm{sum}}}\\
	\mathrm{s.t.~}&\mathrm{C1'}:~{q}_{k}+E_{c,k}+{{\varepsilon }_{k}}f_{k}^{3}T\le{E}_{k},~\forall k,\\ 
	& \mathrm{C7'}:~0\le {{q}_{k}},~\forall k,\\ 
	& \mathrm{C9'}:~\sum\limits_{k=1}^{K}\!\!\left(R_{k}^b\!+Bt_{k}^{a}{{\log }_{2}}\!\left(\!1\!+\!\frac{{{q}_{k}}{{\varpi }_{k}^{*}}}{{{t}_{k}^a\sigma }^2}\!\right)\!\right){{\varphi }_{k}}\!\\
	&~~~~~~~~~~~~\le\!{{f}_{\max}}{{t}_{c}},~\forall k,\\ 
	&\mathrm{C2},~\mathrm{C4},~\mathrm{C5},~\mathrm{C8},
\end{align*}
where $\bm{q}=\{q_k, k\in\mathcal{K}\}$. Nevertheless, the non-convexity of $(\mathbf{P1})$ persists since the objective function and $\mathrm{C9'}$ are still non-convex. To deal with this challenge, a BCD-based optimization framework is  leveraged. In this framework, we divide $(\mathbf{P1})$ into two sub-problems for optimizing $\{\bm{t}, \bm{p}, \bm{f}\}$ and $\{\{\bm{\omega}_{0,k}\}_{k=1}^K, \mathbf\Theta_ {x}\}$, respectively.

\subsection{Optimization of $\{\mathit{t}, \mathit{p}, \mathit{f}\}$ }\label{op}
We first optimize $\{\bm{t}, \bm{p}, \bm{f}\}$ with given $\{\{\bm{\omega}_{0,k}\}_{k=1}^K, \mathbf\Theta_ {x}\}$. ($\mathbf{P1}$) can be reduced as
{
	\begin{align*} 	
		(\mathbf{P2}):&\max_{\bm{t}, \bm{p}, \bm{f}}~~R_{\rm{sum}} \\
		\mathrm{s.t.~}&\mathrm{C1'},~\mathrm{C4},~\mathrm{C5},~\mathrm{C7'},~\mathrm{C9'}.
	\end{align*}
}

By analyzing $\mathrm{C1'}$, it is straightforward to derive that the maximum SCBs is obtained when each WS fully utilizes its harvested energy, which is formulated as 
\begin{equation}
	{{q}_{k}}+{{E}_{c,k}}+{{\varepsilon }_{k}}f_{k}^{3}T={{E}_{k}},~\forall k.
\end{equation}
It is because if there exists any residual energy at each WS, it can be used for further active offloading or local computing\cite{10210080}. Accordingly, ${f_k}$ is written as
\begin{equation}
	{{f}_{k}}={{\left( \frac{{{E}_{k}}-{{q}_{k}}-{{E}_{c,k}}}{{{\varepsilon }_{k}}T} \right)}^{\frac{1}{3}}} .
\end{equation}
Then, $R_k^e$ defined in (\ref{Rek}) can be rewritten as
\begin{equation}
	R_{k}^{e}=\frac{T{{f}_{k}}}{{{\varphi }_{k}}}=\frac{{{T}^{\frac{2}{3}}}}{{{\varphi }_{k}}}{{\left( \frac{{{E}_{k}}-{{q}_{k}}-{{E}_{c,k}}}{{{\varepsilon }_{k}}} \right)}^{\frac{1}{3}}}.
\end{equation}
Then, we can transform ($\mathbf{P2}$) as
\begin{align*} 	
	(\mathbf{P2.1}):&\max_{\bm{t}, \bm{p}}~\sum\limits_{k=1}^{K} \frac{{{T}^{\frac{2}{3}}}}{{{\varphi }_{k}}}{{\left( \frac{{{E}_{k}}-{{q}_{k}}-{{E}_{c,k}}}{{{\varepsilon }_{k}}} \right)}^{\frac{1}{3}}}+\\
	&\sum\limits_{k=1}^{K}\!\!\left(R_k^b+Bt_{k}^{a}{{\log }_{2}}\!\left(\!\!1\!+\!\!\frac{{{q}_{k}}{{\varpi}_{k}^{*}}} {{t}_{k}^a{\sigma }^{2}}\right)\right)\!\! \\
	\mathrm{s.t.}~&\mathrm{C4},~\mathrm{C5},~\mathrm{C7'},~\mathrm{C9'}.
\end{align*}

The convexity of ($\mathbf{P2.1}$) is readily demonstrated, and its optimal solution can be obtained by utilizing the CVX \cite{grant2014cvx}.

\subsection{Optimization of $\{\{{\omega}_{0,k}\}_{k=1}^K, \mathbf\Theta_{x}\}$ }
Based on the solution derived in Section \ref{op}, we then optimize $\{\bm{\omega}_{0,k}\}_{k=1}^K$ and $\mathbf\Theta_{x}$ by solving the following sub-problem with the fixed $\{\bm{t}, \bm{p}, \bm{f}\}$.
{
	\begin{align*} 	
		(\mathbf{P3}):&\max_{\{\bm{\omega}_{0,k}\}_{k=1}^K,\mathbf\Theta_x}~{{R}_{\rm{sum}}}\\
		\mathrm{s.t.}~&\mathrm{C1'},~\mathrm{C2},~\mathrm{C8},~\mathrm{C9'}.
	\end{align*}
}

We first tackle the non-convexity caused by $E_k$ in $\mathrm{C1'}$. In particular, auxiliary variables $z_{k,i}$ and $c_k$ are introduced \cite{SI}, which satisfy  
\begin{equation}\label{EH}
	{{z}_{k,i}}\ge\exp\left(-a_k\left(|{\bm{g}_k^H}\bm{\omega}_{0,i}|^2-b_k \right) \right),
\end{equation}
\begin{equation}\label{EH2}
	\begin{aligned}
		{{c}_{k}}\ge\exp\left(-a_k\left((1-\rho)|{\bm{g}_k^H}\bm{\omega}_{0,k}|^2-b_k \right) \right). 
	\end{aligned}
\end{equation}
Then, by applying the SCA method, the first-order Taylor expansion of $\frac{1}{1+{{z}_{k,i}}}$ is
\begin{equation}\label{z}
	\begin{aligned}
		\frac{1}{1+{{z}_{k,i}}}&\ge \frac{1}{1+z_{k,i}^{(s)}}-\frac{1}{{{\left( 1+z_{k,i}^{(s)} \right)}^{2}}}\left( {{z}_{k,i}}-z_{k,i}^{(s)} \right)\\
		&\triangleq \Gamma\left(\!{{z}_{k,i}},z_{k,i}^{(s)}\!\right)\!,
	\end{aligned}
\end{equation}where $z_{k,i}^{(s)}$ is a feasible point of ${{z}_{k,i}}$ in the $s$-th iteration of implementing the SCA method, and $s$ is used to refer to the iteration number of the SCA method in the following two sub-problems\footnotemark[4].\footnotetext[4]{In Sections \ref{oeb} and \ref{ocm}, we replace $s$ here with $s_3$ and $s_4$ to represent different processes of the SCA iterations.} Similarly, the lower-bound of $\frac{1}{1+{{c}_{k}}}$ is approximated as
\begin{equation}\label{cc}
	\begin{aligned}
		\frac{1}{1+{{c}_{k}}}&\ge \frac{1}{1+c_{k}^{(s)}}-\frac{1}{{{\left( 1+c_{k}^{(s)} \right)}^{2}}}\left( {{c}_{k}}-c_{k}^{(s)} \right)\\
		&\triangleq \Gamma\left(\!{{c}_{k}},c_{k}^{(s)}\!\right)\!,
	\end{aligned}
\end{equation}
where $c_{k}^{(s)}$ is a feasible point of ${{c}_{k}}$ in the $s$-th iteration.

Accordingly, $E_k$ can be approximated as 
\begin{equation}
	\begin{aligned}
		E_k\approx &t_{k}^{b}\left( \Gamma\left( {{c}_{k}},c_{k}^{(s)} \right){{X}_{k}}-{{Y}_{k}}\right)+\\
		&\sum_{i=1,i\neq{k}}^{K}{t_{i}^{b}}\left(\Gamma\left( {{z}_{k,i}},z_{k,i}^{(s)} \right){{X}_{k}}-{{Y}_{k}} \!\right). 
	\end{aligned}
\end{equation}

To ensure the convergence speed and reduce the complexity, we further introduce a variable ${{\alpha }_{k}}\in{[0,1]}$, which denotes the energy allocation between active offloading and local computing in the active offloading phase for the WS-$k$ \cite{10210080}. Using ${{\alpha }_{k}}$, we have 
\begin{equation}\label{q_k}
	q_k={{\alpha }_{k}}\left( {{E}_{k}}-{{E}_{c,k}} \right)
\end{equation}representing the energy allocated for active offloading, and
\begin{equation}
	{{\varepsilon }_{k}}f_{k}^{3}T=\left( 1-{{\alpha }_{k}} \right)\left( {{E}_{k}}-{{E}_{c,k}}\right)
\end{equation}representing the energy used for local computing. 
Then, ${{\alpha }_{k}}$ and ${{f}_{k}}$ can be expressed as
\begin{equation}
	{{\alpha }_{k}}=\frac{q_k}{ {{E}_{k}}-{{E}_{c,k}}}
\end{equation}
and\begin{equation}\label{f_k}
	{{f}_{k}}=\left({\frac{\left( 1-{{\alpha }_{k}} \right)\left( {{E}_{k}}-{{E}_{c,k}}\right)}{{{\varepsilon }_{k}}T}}\right)^{\frac{1}{3}}, 
\end{equation}
respectively. Substituting (\ref{q_k}) and (\ref{f_k}) into (\ref{Rka}) and (\ref{Rek}), the computational bits achieved in the active offloading phase at the WS-$k$ are rewritten as
\begin{equation}
	R_{k}^{a}= Bt_{k}^{a}{{\log }_{2}}\left( 1+\frac{{{\alpha }_{k}}\left( {{E}_{k}}-{{E}_{c,k}} \right){{\varpi }_{k}^{*}}}{t_{k}^{a}\sigma _{{}}^{2}} \right),
\end{equation}
and the computational bits achieved by local computing at the WS-$k$ are reformulated as
\begin{equation}
	R_{k}^{e}=\frac{{{T}^{\frac{2}{3}}}}{{{\varphi }_{k}}}{{\left( \frac{\left( 1-{{\alpha }_{k}} \right)\left( {{E}_{k}}-{{E}_{c,k}} \right)}{{{\varepsilon }_{k}}} \right)}^{\frac{1}{3}}}.
\end{equation}

To simplify the notations, we let
\begin{equation}
	{{R}_{\rm{total}}}=\sum\limits_{k=1}^{K}\left(R_{k}^{e}+ R_{k}^{a}\right).
\end{equation}
Thus, ($\mathbf{P3}$) can be transformed into
{
	\begin{align*} 	
		(\mathbf{P3.1}):&\max_{\{\bm{\omega}_{0,k}\}_{k=1}^K,\mathbf\Theta_x,\bm{z},\bm{c}}~\sum\limits_{k=1}^{K}R_k^b+{{R}_{\rm{total}}}\\
		\mathrm{s.t.}~&\mathrm{C10}:~\sum\limits_{k=1}^{K}\left(R_k^b+Bt_{k}^{a}\right.\\
		&~~~~~~~~~\left.{{\log }_{2}}\!\left(\!\!1\!+\!\frac{{{\alpha }_{k}}\left( {{E}_{k}}-{{E}_{c,k}} \right){{\varpi }_{k}^{*}}}{t_{k}^{a}\sigma _{{}}^{2}} \right)\!\!\right){{\varphi }_{k}}\\
		&~~~~~~~~~\le\!{{f}_{\max}}{{t}_{c}}, \\  
		&\mathrm{C2},~\mathrm{C8},~(\ref{EH}),~(\ref{EH2}),
	\end{align*}
}where $\bm{z}=\{z_{k,i}, k\in\mathcal{K}, i\in\mathcal{K}/k\}$, $\bm{c}=\{c_{k}, k\in\mathcal{K}\}$. Due to the non-convexity of ($\mathbf{P3.1}$), we further split $(\mathbf{P3.1})$ into two sub-problems of optimizing $\{\{\bm{\omega}_{0,k}\}_{k=1}^K,\bm{z},\bm{c}\}$ and $\{\mathbf\Theta_x,\bm{z},\bm{c}\}$, respectively.

\subsubsection{Optimizing energy beamforming}\label{oeb}
Given $\{\bm{t}, \bm{p}, \bm{f}, \mathbf\Theta_x\}$, $(\mathbf{P3.1})$ can be written as 
	{
		\begin{align*} 	
			(\mathbf{P3.2}):&\max_{\{\bm{\omega}_{0,k}\}_{k=1}^K,\bm{z},\bm{c}}~\sum\limits_{k=1}^{K}R_k^b+{{R}_{\rm{total}}}\\
			\mathrm{s.t.}~&\mathrm{C2},~\mathrm{C10},~(\ref{EH}),~(\ref{EH2}).
		\end{align*}
	}
	To handle the non-convexity of $(\mathbf{P3.2})$, we first introduce ${\bm{W}_{k}}\triangleq {\bm{\omega}_{0,k}}\bm{\omega}_{0,k}^{H}\in {{\mathbb{C}}^{\widetilde{N}\times \widetilde{N}}}$, which satisfies ${\bm{W}_{k}}{\succeq }0$ and $\mathrm{rank}\left( {\bm{W}_{k}} \right)=1$. In addition, we let ${\bm{H}_{k}}\triangleq {\bm{h}_{k}}{\bm{g}_{k}^H}\in {{\mathbb{C}}^{\widetilde{N}\times \widetilde{N}}}$, ${\bm{G}_{k}}\triangleq {\bm{g}_{k}}{\bm{g}_{k}^H}\in {{\mathbb{C}}^{\widetilde{N}\times \widetilde{N}}}$ and ${\bm{Z}_{k}}\triangleq {\bm{\omega}_{1,k}}{\bm{\omega}_{1,k}^H}\in {{\mathbb{C}}^{\widetilde{N}\times \widetilde{N}}}$. Then, we rewrite $(\ref{EH}),~(\ref{EH2})$ and $R_k^b$ in $(\ref{Rkb})$ as
	\begin{equation}\label{EH3}
		{{\check{z}}_{k,i}}\ge\exp \left( -{{a}_{k}}\left(  \mathrm{tr}\left( {{{{\bm{G}}}}_{k}}{{{\bm{W}}}_{i}} \right)-\!{{b}_{k}}\! \right)\right),
	\end{equation}
	\begin{equation}\label{EH4}
		\begin{aligned}
			{\check{c}_{k}}\ge&\exp \left(\!-{{a}_{k}}\!\left(\left(1-\rho\right)\mathrm{tr}\left({{{{\bm{G}}}}_{k}}{{{\bm{W}}}_{k}} \right)-{{b}_{k}}\right)\right)
		\end{aligned}
	\end{equation}
	and
	\begin{equation}\label{Rkb'}
		{\check{R}_{k}^b}=Bt_{k}^{b}{{\log }_{2}}\left(\!\!1\!\!+\dfrac{\xi \rho {\rm{tr}}\left(\bm{W}_{k}\bm{H}_{k}^{H}\bm{Z}_k\bm{H}_{k}\right)}{\beta \gamma {\rm{tr}}\left({\bm{W}_{k}}\bm{b}_{{}}^{H}\bm{Z}_k\bm{b}\right)+\sigma^{2}{\rm{tr}}\left(\bm{Z}_k\right)}\right),
	\end{equation}respectively. However, ${\check{R}_{k}^b}$ in (\ref{Rkb'}) is still non-convex due to  the fractional form. To handle this, we use the SCA method to derive its lower bound, which is expressed as
	\begin{equation}
		\begin{aligned}
			{\check{R}_{k}^b}&\ge Bt_{k}^{b}\left({{\log }_{2}}\left(\xi \rho {\rm{tr}}\left(\bm{W}_{k}\bm{H}_{k}^{H}\bm{Z}_k\bm{H}_{k}\right)\right.\right.\\
			&+\left.\left.\beta \gamma{\rm{tr}}\left({\bm{W}_{k}}\bm{b}_{{}}^{H}\bm{Z}_k\bm{b}\right)+\sigma^{2}{\rm{tr}}\left(\bm{Z}_k\right)\right)\right.\\
			&\left.-{{\log }_{2}}\left(\beta \gamma {\rm{tr}}\left(\bm{W}_{k}^{\left(s_3\right)}\bm{b}_{{}}^{H}\bm{Z}_k\bm{b}\right)+\sigma^{2}{\rm{tr}}\left(\bm{Z}_k\right)\right)\right.\\
			&\left.-\frac{\beta \gamma {\rm{tr}}\left(\left(\bm{W}_{k}-\bm{W}_{k}^{\left(s_3\right)}\right)\bm{b}_{{}}^{H}\bm{Z}_k\bm{b}\right)}{{\rm{ln}(2)}\left(\beta \gamma {\rm{tr}}\left(\bm{W}_{k}^{\left(s_3\right)}\bm{b}_{{}}^{H}\bm{Z}_k\bm{b}\right)+\sigma^{2}{\rm{tr}}\left(\bm{Z}_k\right)\right)}\right)\\
			&\triangleq{\widetilde{R}_{k}^b},
		\end{aligned}
	\end{equation}
	where $\bm{W}_{k}^{\left(s_3\right)}$ is a feasible point of $\bm{W}_{k}$ in the $s_3$-th iteration of implementing the SCA method. It can be found that ${\widetilde{R}_{k}^b}$ is a convex function with respect to $\bm{W}_{k}$. By updating the objective function and $\mathrm{C10}$ with $\widetilde{R}_{k}^b$ and relaxing the constraint $\mathrm{rank}\left( {\bm{W}_{k}} \right)=1$, $(\mathbf{P3.2})$ can be converted into the following convex optimization problem
	
	{
		\begin{align*} 	
			(\mathbf{P3.3}):&\max_{\{\bm{\omega}_{0,k}\}_{k=1}^K,\check{\bm{{z}}},\check{\bm{{c}}}}~\sum\limits_{k=1}^{K}\widetilde{R}_k^b+{{R}_{\rm{total}}}\\
			\mathrm{s.t.}~&\mathrm{C10'}:~\!\!\sum\limits_{k=1}^{K}\left(\widetilde{R}_k^b+Bt_{k}^{a}\right.\\
			&~~~~~~~~~~~\left.{{\log }_{2}}\left(\!1\!+\!\frac{{{\alpha }_{k}}\left( {{E}_{k}}-{{E}_{c,k}} \right){{\varpi }_{k}^{*}}}{t_{k}^{a}\sigma _{{}}^{2}} \right)\!\!\right){{\varphi }_{k}}\\
			&~~~~~~~~~~~\le{{f}_{\max}}{{t}_{c}},\\ 
			&\mathrm{C2},~(\ref{EH3}),~(\ref{EH4}),
		\end{align*}
	}
	where $\check{\bm{{z}}}=\{\check{z}_{k,i}, k\in\mathcal{K}, i\in\mathcal{K}/k\}$, $\check{\bm{{c}}}=\{c_{k}, k\in\mathcal{K}\}$.
	
	$(\mathbf{P3.3})$ is a convex optimization problem and its optimal solution can be obtained by using the CVX. According to \cite{10153701}, it is proved that $\bm{W}_{k}$ obtained from $(\mathbf{P3.3})$ is a rank-one matrix, which ensures the optimality of solving $(\mathbf{P3.3})$ after removing the rank-one constraint. By iteratively solving $(\mathbf{P3.3})$ until the objective function converges, the optimal transmit beamforming matrix $\bm{W}_{k}^*$ is finally obtained. Then, the optimal transmit beamforming vector, denoted by $\bm{\omega}_{0,k}^*$, can be obtained using singular value decomposition (SVD) of $\bm{W}_{k}^*$.

\subsubsection{Optimizing the coefficient matrices of STAR-RIS in the passive offloading phase} \label{ocm}
Given $\bm{t}, \bm{p}, \bm{f}$, and $\{\bm{\omega}_{0,k}\}_{k=1}^K$, we continue to optimize the coefficient matrices of the STAR-RIS in the passive offfloading phase by solving the following problem

{
	\begin{align*} 	
		(\mathbf{P3.4}):&\max_{\mathbf\Theta_x,\bm{z},\bm{c}}~\sum\limits_{k=1}^{K}R_k^b+{{R}_{\rm{total}}}\\
		\mathrm{s.t.}~&\mathrm{C8},~\mathrm{C10},~(\ref{EH}),~(\ref{EH2}).
	\end{align*}
}

Due to the quartic form of ${\mathbf\Theta_{x}}$, ($\mathbf{P3.4}$) is also a non-convex problem. We exploit SDR and SCA techniques to address this challenge. To be specific, we set ${\bm{a}_{k}}=\mathrm{diag}\left( \bm{\omega}_{1,k}^{H}\bm{H}_{\textit{s,c}}^H \right)\bm{h}_{\textit{s,x}}\in {{\mathbb{C}}^{M\times1}}$, ${{b}_{k}}=\bm{\omega}_{1,k}^{H}{\bm{h}_{c,x}}$, ${\bm{c}_{k}}=\mathrm{diag}\left( \bm{g}_{s,x}^{H} \right){\bm{G}_{p,s}}{\bm{\omega}_{0,k}}\in {{\mathbb{C}}^{M\times 1}}$, and ${{d}_{k}}=\bm{g}_{p,x}^{H}{\bm{\omega}_{0,k}}$. Let $\bm{v}_{x}^{{}}={{\left[ \sqrt{\beta _{1}^{x}}{{e}^{j\theta _{1}^{x}}},\sqrt{\beta _{2}^{x}}{{e}^{j\theta _{2}^{x}}},\ldots ,\sqrt{\beta _{M}^{x}}{{e}^{j\theta _{M}^{x}}} \right]}^{H}}$ and $\bm{\bar{v}}_{x}=\left [{\bm{v}_{x}^H,1}\right]^H$ with $x\in\{r,t\}$. Accordingly, $|\bm{\omega}_{1,k}^H\bm{h}_k\bm{g}_k^H\bm{\omega}_{0,k}|^2$ in the objective function and $\mathrm{C10}$ is reformulated as
\begin{equation} \label{eq2}
	\begin{aligned}
		\varPsi\left( \bm{\bar{v}}_{x} \right)=&\bm{\bar{v}}_{x}^{H}{\bm{A}_{k}}\bm{\bar{v}}_{x}^{{}}\bm{\bar{v}}_{x}^{H}{\bm{C}_{k}}\bm{\bar{v}}_{x}^{{}}+{{\left| {{d}_{k}} \right|}^{2}}\bm{\bar{v}}_{x}^{H}{\bm{A}_{k}}\bm{\bar{v}}_{x}^{{}}\\
		&+{{\left| {{b}_{k}} \right|}^{2}}\bm{\bar{v}}_{x}^{H}{\bm{C}_{k}}\bm{\bar{v}}_{x}^{{}}+{{\left| {{b}_{k}} \right|}^{2}}{{\left| {{d}_{k}} \right|}^{2}},
	\end{aligned}
\end{equation}
where
$\bm{A}_{k}={
	\left[ \begin{array}{cc}
		{\bm{{a}}_{k}}\bm{a} _{k}^{H} & {b} _{k}^H\bm{a}_{k} \\
		{b} _{k}\bm{a}_{k}^H & 0 \\
	\end{array} 
	\right ]}$, $\bm{C}_{k}={
	\left[ \begin{array}{cc}
		{\bm{{c}}_{k}}\bm{c} _{k}^{H} & {d} _{k}^H\bm{c}_{k} \\
		{d} _{k}\bm{c}_{k}^H & 0  \\
	\end{array} 
	\right ]}$. Since (\ref{eq2}) is a quartic polynomial of $\bm{\bar{v}}_{x}$, it is challenging to handle directly. Similar to \cite{9611544}, we leverage the SCA method to handle $\varPsi\left( \bm{\bar{v}}_{x} \right)$. By applying the SCA method, the lower bound of a non-concave function with respect to $\bm{\chi}$, denoted by $\psi\left(\bm{\chi}\right)$, can be expressed as
\begin{equation} \label{eq3}
	\begin{aligned}
		\psi\left( \bm{\chi} \right)\ge&\psi\left( {\bm{\chi}_{0}} \right)+\operatorname{Re}\left\{ \nabla \psi{{\left( {\bm{\chi}_{0}} \right)}^{H}}\left( \bm{\chi}-{\bm{\chi}_{0}} \right) \right\}\\
		&-\frac{l}{2}{{\left\| \bm{\chi}-{\bm{\chi}_{0}} \right\|}^{2}},
	\end{aligned}
\end{equation}where the first order Taylor expansion is utilized, ${\bm{\chi}_{0}}$ is a feasible point of ${\bm{\chi}}$, $\nabla $ represents the gradient operator, and $l$ represents the maximum curvature of $\psi\left(\bm{\chi}\right)$. Accordingly, by utilizing this method, (\ref{eq2}) undergoes an approximation transformation resulting in a concave function as shown below
\begin{equation}
	\begin{aligned}
		\varPsi\left( \bm{\bar{v}}_{x} \right)\ge& {(\bm{\bar{v}}_{x}^{(s_4)})^{H}}{\bm{A}_{k}}{\bm{\bar{v}}_{x}^{(s_4)}}{(\bm{\bar{v}}_{x}^{(s_4)})^{H}}{\bm{C}_{k}}{\bm{\bar{v}}_{x}^{(s_4)}}+{{\left| {{d}_{k}} \right|}^{2}} \\ 
		&{(\bm{\bar{v}}_{x}^{(s_4)})^{H}}{\bm{A}_{k}}{\bm{\bar{v}}_{x}^{(s_4)}}+{{\left| {{b}_{k}} \right|}^{2}}{(\bm{\bar{v}}_{x}^{(s_4)})^{H}}{\bm{C}_{k}}{\bm{\bar{v}}_{x}^{(s_4)}}\\ 
		&+\!\!{{\left| {{d}_{k}} \right|}^{2}}{{\left| {{b}_{k}} \right|}^{2}}\!+\!{(\bm{\bar{v}}_{x}^{(s_4)})^{H}}{\bm{T}_{x}}\left(\!\bm{\bar{v}}_{x}\!-\!{\bm{\bar{v}}_{x}^{(s_4)}}\!\right)+\\
		&{{\left(\!\bm{\bar{v}}_{x}\!-\!{\bm{\bar{v}}_{x}^{(s_4)}}\!\right)}^{H}}{\bm{T}_{x}}{\bm{\bar{v}}_{x}^{(s_4)}}-\frac{l}{2}\!\left(\!\! \bm{\bar{v}}_{x}^{H}\bm{\bar{v}}_{x}\!-\!\bm{\bar{v}}_{x}^{H}{\bm{\bar{v}}_{x}^{(s_4)}}\!\right.\\
		&\left.-\!{(\bm{\bar{v}}_{x}^{(s_4)})^{H}}\bm{\bar{v}}_{x}\!+\!{{\left\| \bm{\bar{v}}_{x}^{(s_4)}\right\|}^{2}}\! \right) \\ 
		=&-\frac{l}{2}\left( \bm{\bar{v}}_{x}^{H}\bm{\bar{v}}_{x}-\bm{\bar{v}}_{x}^{H}{\bm{\bar{v}}_{x}^{(s_4)}}-{(\bm{\bar{v}}_{x}^{(s_4)})^{H}}\bm{\bar{v}}_{x}+\right. \\ 
		&\left.{{\left\| {\bm{\bar{v}}_{x}^{(s_4)}} \right\|}^{2}} \right)+{(\bm{\bar{v}}_{x}^{(s_4)})^{H}}{\bm{T}_{x}}\bm{\bar{v}}_{x}+\bm{\bar{v}}_{x}^{H}{\bm{T}_{x}}{\bm{\bar{v}}_{x}^{(s_4)}}\\ 
		&+\widehat{c} \\
		=&-\frac{l}{2}\!\left(\! \bm{\bar{v}}_{x}^{H}I\bm{\bar{v}}_{x}\!+\!\bm{\bar{v}}_{x}^{H}\left(\!-\frac{2}{l}{\bm{T}_{x}}{\bm{\bar{v}}_{x}^{(s_4)}}-I{\bm{\bar{v}}_{x}^{(s_4)}} \!\right) \right. \\ 
		&\left.+{{\left( -\frac{2}{l}{\bm{T}_{x}}{\bm{\bar{v}}_{x}^{(s_4)}}-I{\bm{\bar{v}}_{x}^{(s_4)}} \right)}^{H}}\bm{\bar{v}}_{x} \right)+c,
	\end{aligned}
\end{equation}
where ${\bm{\bar{v}}_{x}^{(s_4)}}$ represents a feasible point of $\bm{\bar{v}}_{x}$ in the $s_4$-th iteration of the SCA method, ${\bm{T}_{x}}\triangleq{\bm{A}_{k}}{{\bm{\bar{v}}_{x}^{(s_4)}}}{({\bm{\bar{v}}_{x}^{(s_4)}})}^{H}{\bm{C}_{k}}+{\bm{C}_{k}}{{\bm{\bar{v}}_{x}^{(s_4)}}}{({\bm{\bar{v}}_{x}^{(s_4)}})}^{H}{\bm{A}_{k}}+{{\left| {{b}_{k}} \right|}^{2}}{\bm{C}_{k}}+{{\left| {{d}_{k}} \right|}^{2}}{\bm{A}_{k}}$, $\widehat{c}={({\bm{\bar{v}}_{x}^{(s_4)}})}^{H}\!\!\left({\bm{A}_{k}}{\bm{\bar{v}}_{x}^{(s_4)}}{({\bm{\bar{v}}_{x}^{(s_4)}})}^{H}\bm{C}_{k}+{{\left| {{d}_{k}} \right|}^{2}}{\bm{A}_{k}}+{{\left| {{b}_{k}} \right|}^{2}}{\bm{C}_{k}}-\right.\\
\left.2{\bm{T}_{x}}\right){\bm{\bar{v}}_{x}^{(s_4)}}+{{\left| {{d}_{k}} \right|}^{2}}{{\left| {{b}_{k}} \right|}^{2}}$, and $c=\widehat{c}-\frac{l}{2}\Vert{\bm{\bar{v}}_{x}^{(s_4)}}\Vert^2$ represents all the terms unrelated with $\bm{\bar{v}}_{x}$. Let $\bm{\amalg}=-\frac{2}{l}{\bm{T}_{x}}{\bm{\bar{v}}_{x}^{(s_4)}}-\bm{I}{\bm{\bar{v}}_{x}^{(s_4)}}$,
\begin{equation}\label{U}
	\bm{U}_{x}=-\frac{l}{2}\!\!\times\!\!{
		\left[\!\!\begin{array}{cc}
			\bm{I} & \bm{\amalg}  \\
			\bm{\amalg}^{H} & 0 
		\end{array}\!\! 
		\right]},
\end{equation}
$\bm{\bar{\bar{v}}}_{x}={
	\left[ \begin{array}{c}
		\bm{\bar{v}}_{x}\\
		1  \\
	\end{array} 
	\right ]}$, and ${{\bm{\bar{\bar{V}}}}_{x}}=\bm{\bar{\bar{v}}}_{x}\bm{\bar{\bar{v}}}_{x}^{H}$, where ${\rm{rank}}\left( \bm{\bar{\bar{V}}}_{x} \right)=1$. Then, $|\bm{\omega}_{1,k}^H\bm{h}_k\bm{g}_k^H\bm{\omega}_{0,k}|^2$ is rewritten as
\begin{equation}
	|\bm{\omega}_{1,k}^H\bm{h}_k\bm{g}_k^H\bm{\omega}_{0,k}|^2={\rm{tr}}\left(\bm{U}_{x}\bar{\bar{\bm{V}}}_{x} \right)+c.
\end{equation}

Similarly, the quadratic polynomial of $\bm{\bar{v}}_{x}$ in $E_k$, i.e., $|{\bm{g}_k^H}\bm{\omega}_{0,k}|^2$, is reformulated as
\begin{equation}
	|{\bm{g}_k^H}\bm{\omega}_{0,k}|^2=\bm{\bar{v}}_{x}^{H}{\bm{C}_{k}}\bm{\bar{v}}_{x}+{{\left|d_{k}\right|}^{2}}={\rm{tr}}\left(\bar{\bm{C}}_{k}\bar{\bar{\bm{V}}}_{x} \right)+{{\left|d_{k}\right|}^{2}},
\end{equation}
where $\bm{\bar{C}}_{k}={
	\left[ \begin{array}{cc}
		\bm{{C}}_{k}& 0 \\ 
		0& 0 \\ 
	\end{array} 
	\right]}\in {{\mathbb{C}}^{\left( M+2 \right)\left( M+2 \right)}}$.
Then, (\ref{EH}) and (\ref{EH2}) can be rewritten as
\begin{equation}\label{EH5}
	{\hat{z}_{k,i}}\ge\exp \left( -{{a}_{k}}\left( \left( \mathrm{tr}\left( {{{\bar{\bm{C}}}}_{k,i}}{{{\bar{\bar{\bm{V}}}}}_{x}} \right)\!+\!{{\left| d_{k}^{{}} \right|}^{2}} \right)\!\!-\!{{b}_{k}}\! \right)\!\right)\!\!,
\end{equation}
\begin{equation}\label{EH6}
	\begin{aligned}
		{\hat{c}_{k}}\ge&\exp \left(\!-{{a}_{k}}\!\left(\left(1-\rho\right)\left(\mathrm{tr}\left( {{{\bar{\bm{C}}}}_{k}}{{{\bar{\bar{\bm{V}}}}}_{x}} \right)+{{\left| d_{k}^{{}} \right|}^{2}} \right)\right.\right.\\
		&\left.\left.-{{b}_{k}}\right)\right), 
	\end{aligned}
\end{equation}
respectively. We then let

\begin{align*}
	{\hat{R}_{k}^b}=B t_{k}^{b}{{\log }_{2}}\left(\!\!1\!\!+\!\!\frac{\xi \rho {({\rm{tr}}\left(\bm{U}_{x}\bar{\bar{\bm{V}}}_{x} \right)+c)}}{{\beta\gamma|\bm{\omega}_{1,k}^H\bm{b}\bm{\omega}_{0,k}|^{2}+\sigma^2\left\|\bm{\omega}_{1,k}\right\|^2}} \!\!\right).
\end{align*}Thus, ($\mathbf{P3.4}$) can be transformed into
\begin{align*} 	
	(\mathbf{P3.5}):&\max_{{\bar{\bar{\bm{{V}}}}_x}, \bm{\beta }^{x},\hat{\bm{z}},\hat{\bm{c}}}~\sum\limits_{k=1}^{K}{\hat{R}_{k}^b}+{{R}_{\rm{total}}} \\
	\mathrm{s.t.}~&\mathrm{C10''}:~\!\!\sum\limits_{k=1}^{K}\left({\hat{R}_{k}^b}+Bt_{k}^{a}\right.\\
	&~~~~~~~~~~~\left.{{\log }_{2}}\left(\!1\!+\!\frac{{{\alpha }_{k}}\left( {{E}_{k}}-{{E}_{c,k}} \right){{\varpi }_{k}^{*}}}{t_{k}^{a}\sigma _{{}}^{2}} \right)\!\!\right){{\varphi }_{k}}\\
	&~~~~~~~~~~~\le{{f}_{\max}}{{t}_{c}},\\ 
	&\mathrm{C11}:~\mathrm{rank}\left( {{\bar{\bar{\bm{{V}}}}}_{x}} \right)=1, x\in \left\{ r,t \right\}, \\
	&\mathrm{C12}:~\mathrm{diag}\left( {{\bar{\bar{\bm{{V}}}}}_{x}} \right)={{\left[ {\bm{\beta }^{x}},1,1 \right]}^{T}}, x\in \left\{ r,t \right\},\\ 
	&\mathrm{C8},~(\ref{EH5}),~(\ref{EH6}),
\end{align*}where ${\bm{\beta }^{x}}\triangleq {{\left[ \beta _{1}^{x},\beta _{2}^{x},...,\beta _{M}^{x} \right]}^{T}}$, $\hat{\bm{{z}}}=\{\hat{z}_{k,i}, k\in\mathcal{K}, i\in\mathcal{K}/k\}$, and $\hat{\bm{{c}}}=\{c_{k}, k\in\mathcal{K}\}$. 

We then proceed to deal with the non-convexity caused by $\mathrm{C11}$. According to \cite{9483903}, $\mathrm{C11}$ can be converted into a difference form as
\begin{equation}
	{{\left\| {{{\bar{\bar{\bm{V}}}}}_{x}} \right\|}_{*}}-{{\left\| {{{\bar{\bar{\bm{V}}}}}_{x}} \right\|}_{2}}=0,~x\in \left\{ r,t \right\},
\end{equation}where ${{\left\| {{{\bar{\bar{\bm{V}}}}}_{x}} \right\|}_{*}}={\sum_{i}{\sigma }_{i}}\left( {{{\bar{\bar{\bm{{V}}}}}}_{x}} \right)$ and ${{\left\| {{{\bar{\bar{\bm{V}}}}}_{x}} \right\|}_{2}}={{\sigma }_{1}}\left( {{{\bar{\bar{\bm{{V}}}}}}_{x}} \right)$ denote the nuclear norm and spectral norm, respectively, ${{\sigma }_{i}}\left( {{{\bar{\bar{\bm{V}}}}}_{x}} \right)$ denotes the $i$-th singular value of ${{\bar{\bar{\bm{V}}}}_{x}}$, and ${{\sigma }_{1}}\left( {{{\bar{\bar{\bm{V}}}}}_{x}} \right)$ is the largest singular value of ${{\bar{\bar{\bm{V}}}}_{x}}$. It is notable that ${{\left\| {{{\bar{\bar{\bm{V}}}} }_{x}} \right\|}_{*}}-{{\left\| {{{\bar{\bar{\bm{V}}}}}_{x}} \right\| }_{2}}=0$ holds when the rank of ${{\bar{\bar{\bm{V}}}}_{x}}$ is one. Therefore, in the $s_4$-th iteration, a lower-bound of ${{\left\| {{{\bar{\bar{\bm{V}}}}}_{x}} \right\|}_{2}}$ at the point $\bar{\bar{\bm{V}}}_{x}^{\left(s_4\right)}$ is given by
\begin{equation}\label{VV}
	\begin{aligned}
		{{\left\| {{{\bar{\bar{\bm{V}}}}}_{x}}\right\|}_{2}}\ge&{{\left\| \bar{\bar{\bm{V}}}_{x}^{\left(s_4\right)}\right\|}_{2}}+\mathrm{tr}\left[u\left( \bar{\bar{\bm{V}}}_{x}^{\left(s_4\right)} \right){{\left(u\left(\bar{\bar{\bm{V}}}_{x}^{\left(s_4\right)}\right)\right)}^{H}}\right.\\
		&\left.\left(\bar{\bar{\bm{V}}}_{x}-\bar{\bar{\bm{V}}}_{x}^{\left(s_4\right)}\right)\!\right]\\
		=&\Upsilon\left({{{\bar{\bar{\bm{V}}}}}_{x}},\bar{\bar{\bm{V}}}_{x}^{\left(s_4\right)}\right), 
	\end{aligned}
\end{equation}
where $u\left( \bar{\bar{\bm{V}}}_{x}^{\left(s_4\right)} \right)$  is the eigenvector associated to the largest eigenvalue of $\bar{\bar{\bm{V}}}_{x}^{\left(s_4\right)}$. With the lower bound of ${{\left\| {{{\bar{\bar{\bm{V}}}}}_{x}} \right\|}_{2}}$, the approximate form of $\mathrm{C11}$ is expressed as
\begin{equation}
	{{\left\| {{{\bar{\bar{\bm{V}}}}}_{x}} \right\|}_{*}}-\Upsilon \left( {{{\bar{\bar{\bm{V}}}}}_{x}},\bar{\bar{\bm{V}}}_{x}^{\left(s_4\right)} \right)\le {{\varsigma }_{V}},
\end{equation}
where ${{\varsigma }_{V}}$ is a positive threshold whose value is close to 0. The approximate rank-one constraint can guarantee the following inequalities, i.e.,
\begin{equation}
	\begin{aligned}		
		0\le {{\left\| {{{\bar{\bar{\bm{V}}}}}_{x}} \right\|}_{*}}\!\!-\!\!{{\left\| {{{\bar{\bar{\bm{V}}}}}_{x}} \right\|}_{2}}\le {{\left\| {{{\bar{\bar{\bm{V}}}}}_{x}} \right\|}_{*}}\!\!-\!\!\Upsilon \left( {{{\bar{\bar{\bm{V}}}}}_{x}},\bar{\bar{\bm{V}}}_{x}^{\left(s_4\right)} \right)\le {{\varsigma }_{{V}}}, \nonumber		
	\end{aligned}
\end{equation}
and the value of ${{\varsigma }_{V}}$ can be appropriately set to control the approximate precision of $\mathrm{C11}$.

Consequently, ($\mathbf{P3.5}$) can be rewritten as
\begin{align*} 	
	(\mathbf{P3.6}):&\max_{{\bar{\bar{\bm{{V}}}}_x}, \bm{\beta }^{x},\hat{\bm{z}},\hat{\bm{c}}}~\sum\limits_{k=1}^{K}{\hat{R}_{k}^b}+{{R}_{\rm{total}}} \\
	\mathrm{s.t.}~&\mathrm{C13}:~\mathrm{tr}\left( {{{\bar{\bar{\bm{V}}}}}_{x}} \right)-\Upsilon \left( {{{\bar{\bar{\bm{V}}}}}_{x}},\bar{\bar{\bm{V}}}_{x}^{\left(s_4 \right)} \right)\le {{\varsigma }_{V}}, \\
	&\mathrm{C8},~\mathrm{C10''},~\mathrm{C12},~(\ref{EH5}),~(\ref{EH6}).
\end{align*}

($\mathbf{P3.6}$) is observed to be a convex semidefinite program that can be addressed by utilizing the CVX \cite{grant2014cvx}. The optimal solution ${{\bar{\bar{\bm{{V}}}}_x}^{*}}$ can be obtained by solving problem ($\mathbf{P3.6}$). The optimal STAR-RIS coefficients in the passive offloading phase $\bm{v}_{x}^{{*}}$ can be obtained by implementing the SVD of ${{\bar{\bar{\bm{{V}}}}_x}^{*}}$.

\begin{algorithm} 
	\caption{The proposed algorithm for solving ($\mathbf{P0}$).}
	\label{alg:pro}
	\begin{algorithmic}[1]
		\STATE Initialization:
		$\{\{\bm{\omega}_{0,k}\}_{k=1}^K\}^{(0)}$, $\{\{\bm{\omega}_{2,k}\}_{k=1}^K\}^{(0)}$,  $\{\{\mathbf\Theta_{x,k}\}_{k=1}^K\}^{(0)}$, $\bm{t}^{(0)}$, $\bm{p}^{(0)}$, $\check{\bm{z}}^{(0)}$, $\check{\bm{c}}^{(0)}$, ${{\bm{v}}}_{x}^{(0)}$, convergence tolerance $\epsilon$, BCD iteration index $s_1$, iteration index $s_2$ for optimizing $\{\bm{\omega}_{2,k}\}_{k=1}^K$ and $\{\mathbf\Theta_{x,k}\}_{k=1}^K$, SCA iteration index $s_3$ and $s_4$.
		\FOR{$k=1$ to $K$}
		\STATE $s_1=0$.
		\REPEAT
		\STATE $s_1=s_1+1$, $s_2=0$.
		\REPEAT
		\STATE $s_2=s_2+1$.
		\STATE Calculate $\bm{\omega}_{2,k}^{(s_2)}$ based on (\ref{w}).
		\STATE Calculate $\mathbf\Theta_{x,k}^{(s_2)}$ based on (\ref{c}).
		\STATE Calculate ${ \varpi}_k^{(s_2)}=\frac{{\left|\left(\bm{\omega}_{2,k}^{(s_2)}\right)^{H}\bm{h}_{k,2}^{(s_2)} \right|}^{2}}{\left\|\bm{\omega}_{2,k}^{(s_2)}\right\|^2}$.
		\UNTIL the increase of the value of ${ \varpi_k^{(s_2)}}$ is less than $\epsilon$.
		\STATE Update ${ \varpi_k^{(s_1)}}={ \varpi_k^{(s_2)}}$.
		\STATE Calculate $\bm{\omega}_{1,k}^{(s_1)}$ based on (\ref{w1}).
		\STATE $s_3=0$, $s_4=0$.
		\REPEAT
		\STATE $s_3=s_3+1$.
		\STATE Calculate $\Gamma\left( {\check{z}_{k,i}},\check{z}_{k,i}^{(s_3)} \right)$ and $\Gamma\left( {\check{c}_{k}},\check{c}_{k}^{(s_3)} \right)$ based on (\ref{z}) and (\ref{cc}), respectively.
		\STATE Solve ($\mathbf{P3.3}$) to obtain $\check{z}_{k,i}^{(s_3)}$, $\check{c}_{k}^{(s_3)}$, and $\bm{\omega}_{0,k}^{(s_3)}$.
		\UNTIL the increase of the value of the objective function in ($\mathbf{P3.3}$) is less than $\epsilon$.
		\STATE Update $\bm{\omega}_{0,k}^{(s_1)}=\bm{\omega}_{0,k}^{(s_3)}$.
		\STATE Calculate $\hat{z}_{k,i}^{(0)}$ and $\hat{c}_{k}^{(0)}$ based on (\ref{EH5}) and (\ref{EH6}), respectively.
		\REPEAT
		\STATE $s_4=s_4+1$.
		\STATE Calculate $\bm{U}_{x}$ based on (\ref{U}).
		\STATE Calculate $\Gamma\left( {\hat{z}_{k,i}},\hat{z}_{k,i}^{(s_4)} \right)$, $\Gamma\left( {\hat{c}_{k}},\hat{c}_{k}^{(s_4)} \right)$ and $\Upsilon \left( {{{\bar{\bar{\bm{V}}}}}_{x}},\bar{\bar{\bm{V}}}_{x}^{\left(s_4\right)} \right)$ based on (\ref{z}), (\ref{cc}) and (\ref{VV}), respectively.
		\STATE Solve ($\mathbf{P3.6}$) to obtain $\hat{z}_{k,i}^{(s_4)}$, $\hat{c}_{k}^{(s_4)}$, and $\bar{\bar{\bm{v}}}_{x}^{(s_4)}$.
		\UNTIL the increase of the value of the objective function in ($\mathbf{P3.6}$) is less than $\epsilon$.
		\STATE Update $\bar{\bar{\bm{v}}}_{x}^{(s_1)}=\bar{\bar{\bm{v}}}_{x}^{(s_4)}$.
		\STATE Calculate $\check{z}_{k,i}^{(s_1)}$ and $\check{c}_{k}^{(s_1)}$ based (\ref{EH3}) and (\ref{EH4}), respectively.
		\STATE Solve ($\mathbf{P2.1}$) to obtain ${t}_{k}^{(s_1)}, {p}_{k}^{(s_1)}$ and ${f}_{k}^{(s_1)}$.		
		\UNTIL the increase of the value of the objective function in ($\mathbf{P0}$) is less than $\epsilon$.
		\ENDFOR
		\RETURN {$\{\mathbf\Theta_{x}^*, \bm{t}^*, \bm{p}^*, \bm{f}^*, \{\{\mathbf\Theta_{x,k}\}_{k=1}^K\}^*, \bm{\omega}^{*}$\}}.
		
	\end{algorithmic}
\end{algorithm}

\subsection{Convergence and Complexity Analysis}
The overall procedure for solving ($\mathbf{P0}$) is outlined in Algorithm \ref{alg:pro}. It is evident that the SCBs do not exhibit a decline after each iteration of the BCD-based framework. Therefore, the objective function in ($\mathbf{P0}$) will finally tend to a fixed value, as demonstrated in \cite{10032506}.

According to \cite{10224271, 6891348}, the complexity of solving ($\mathbf{P3}$) is expressed as $\mathcal{O}(\sqrt{(K+2)M+2M}\log(\frac{1}{\epsilon})·(n_1(K+2)M^3+{n_1}^2(K+2)M^2+2n_{1}M+{n_1}^3))$, where $n_1=2M+K+2$. The complexity of solving ($\mathbf{P2.1}$) is expressed as $\mathcal{O}(\sqrt{(K+2)N}\log(\frac{1}{\epsilon})·(n_2(K+2)N^3+{n_2}^2(K+2)N^2+{n_2}^{3}))$, where $n_2=N^2$. Furthermore, the computational complexity of updating $\{\{\bm{\omega}_{1,k}\}_{k=1}^K\}$, $\{\{\bm{\omega}_{2,k}\}_{k=1}^K\}$ and $\{\{\mathbf\Theta_{x,k}\}_{k=1}^K\}$ in the closed-form is $\mathcal{O}(1)$ \cite{10210080}. Thus, the computational complexity of Algorithm \ref{alg:pro} is expressed $\mathcal{O}({I}_{2}({I}_{1}\sqrt{(K+2)M+2M}\log(\frac{1}{\epsilon})·(n(K+2){M}^3+{n}^2(K+2){M}^2+2nM+{n}^3)+\sqrt{(K+2)N}\log(\frac{1}{\epsilon})·(n_2(K+2)N^3+{n_2}^2(K+2)N^2+{n_2}^{3})))$, where $I_1$ is the number of iterations to solve ($\mathbf{P3}$) with the SCA method, and $I_2$ is the number of iterations for the BCD-based framework.

\section{Numerical Results}
\label{IV}

\begin{figure}[t]
	\centering
	\includegraphics[height=3.5cm,width=6.5cm]{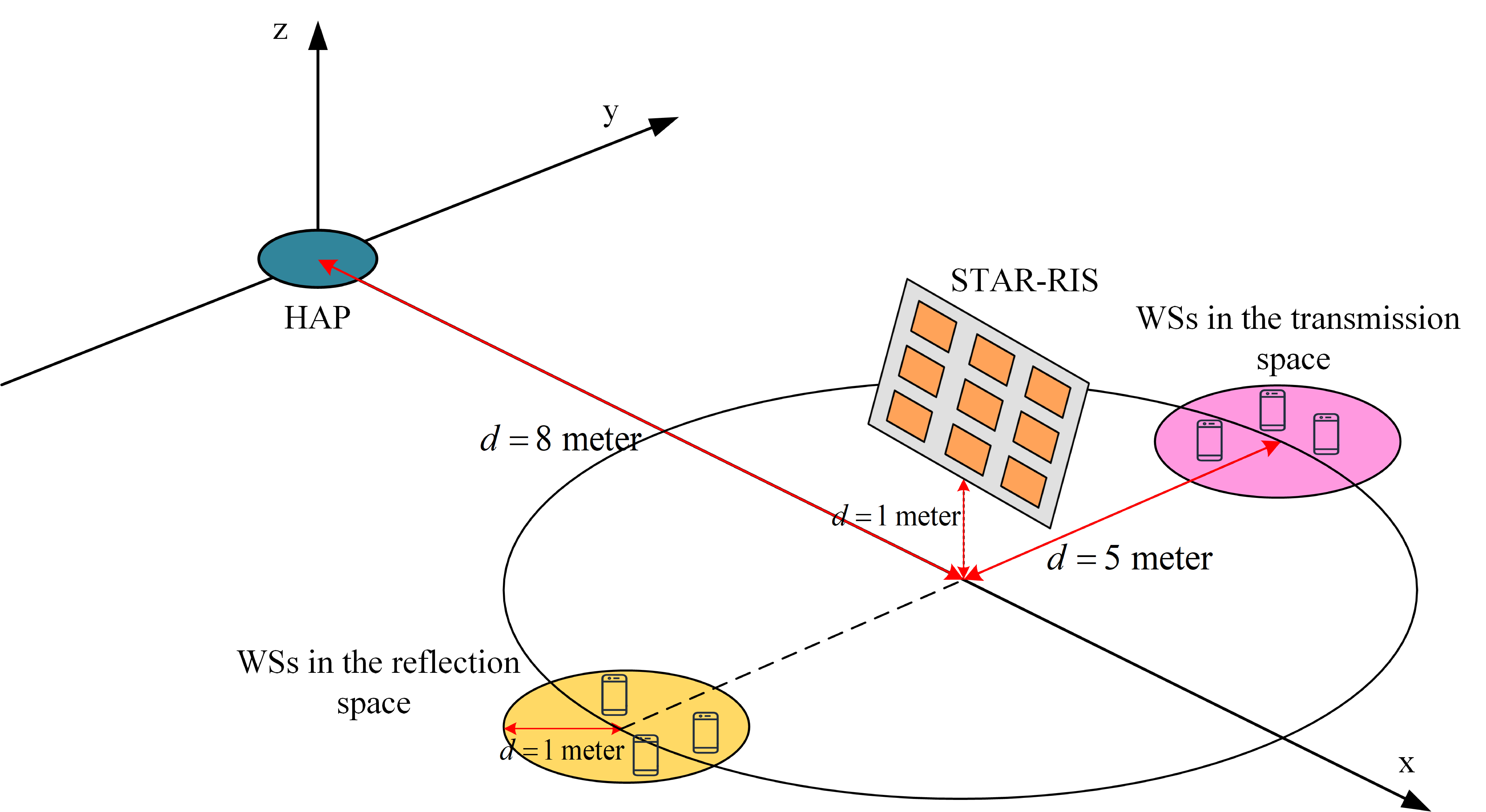}
	\caption{The simulated setup.}
	\label{tuopu}
\end{figure}

In this section, simulation results are presented for evaluating the proposed scheme and algorithm. A computer with 5.2 GHz CPU and 64GB RAM is adopted as the simulation platform. In addition, the CVX solver \cite{grant2014cvx} is adopted in Algorithm \ref{alg:pro}. As depicted in Fig. \ref{tuopu}, a three-dimensional coordinate topology is utilized for simulations. The coordinates of the HAP and the STAR-RIS are (0, 0, 0) meter and (8, 0, 1) meter, respectively. The coordinates of WSs in the reflection and transmission spaces are randomly distributed in two circles, whose centers are set at (8, -5, 0) meter and (8, 5, 0) meter, respectively, and the radius is 1 meter. 

We consider the effects of large-scale fading and small-scale fading on signal transmissions. According to the setting in \cite{9214497}, the large-scale fading is mathematically denoted by $A\left(\frac{d}{d_0}\right)^{-\kappa}$, where $A$ denotes the pass-loss at the reference distance $d_0$ = 1 meter, $d$ represents the distance between the two nodes, and $\kappa$ represents the path-loss factor. To model the small-scale fading for the links connecting to the STAR-RIS, we utilize the widely-adopted Rician fading model. As an example, $\bm{G}_{p,s}$ is expressed as
\begin{equation}
	{\bm{G}_{\textit{p,s}}}=\sqrt{\frac{\eta}{1+\eta}}{\bm{G}_{\textit{p,s}}}^{\text{LoS}}+\sqrt{\frac{1}{1+\eta}}{\bm{G}_{\textit{p,s}}}^{\text{NLoS}},
\end{equation} 	
where $\eta$ is the Rician factor, ${\bm{G}_{\textit{p,s}}}^{\text{LoS}}$ and ${\bm{G}_{\textit{p,s}}}^{\text{NLoS}}$ represent the line-of-sight (LoS) component and non-LoS component, respectively. The small-scale fading characteristics for the other links are defined in a similar manner. Nevertheless, for the links that are not connected to the STAR-RIS, the small-scale fading follows the Rayleigh fading model with $\mathcal{CN}(0,1)$. Without specific statements, we follow the parameters adopted in \cite{9812481} and \cite{SI}, which are summarized in Table \ref{tab1}.

\begin{table}[t]
	\footnotesize
	\renewcommand{\arraystretch}{1.2}
	\caption{Simulation Parameters}
	\label{tab1}
	\centering
	\begin{tabular}{|c|c|c|}
		\hline
		\textbf{Parameters} & \textbf{Notation}  &\textbf{Values} \\
		\hline
		Duration of transmission block & $T$ & 1 s  \\ 
		\hline
		Number of WSs & $K$ & 4 \\
		\hline
		Number of antennas & $N$ & 4 \\
		\hline
		System bandwidth & $B$ & 100 kHz \\
		\hline
		Noise power & $\sigma^2$ & -50 dBm\\
		\hline
		Number of STAR-RIS elements & $M$ & 30  \\
		\hline
		Number of cycles at the WS-$k$ & $\varphi_k$ & 1000 cycles/bit \\
		\hline
		Path-loss factor from HAP to WSs & $\kappa_1$ &3\\
		\hline
		Path-loss factor from STAR-RIS to WSs & $\kappa_2$ &2 \\
		\hline
		Path-loss factor from HAP to STAR-RIS & $\kappa_3$ &2.2 \\
		\hline
		Pass-loss at $d_0$ = 1 m & $A$ & -10 dB\\
		\hline
		Maximum saturation power of EH circuit & $P_{\rm{sat}}$&24 mW\\
		\hline
		EH circuit parameter&$a_k$ &150\\
		\hline
		EH circuit parameter& $b_k$ & 0.0022\\
		\hline
		Rician factor & $\eta$ & 3 dB\\
		\hline
		Performance gap & $\xi$ & 0.0316 \\
		\hline
		Maximum CPU frequency&$f_{\max}$ &$3 × 10^9$ Hz \\
		\hline
		Maximum transmit power at HAP& $P_{t}$ & 1 W\\
		\hline
		Effective capacitance coefficient & $\varepsilon_k$ &${10}^{-26}$\\
		\hline
		Tolerant threshold & ${{\varsigma}_{V}}$ & $10^{-4}$ \\
		\hline
		Maximum curvature & $l$ & $2.5 × 10^{-16}$ \\
		\hline
		Circuit energy consumption  & $E_{c,k}$& $10^{-10}$ J\\
		\hline
		Power reflection coefficient of WSs &$\rho$ & 0.8\\ 
		\hline
	\end{tabular}
\end{table}
To ensure an extensive performance comparison, the following representative benchmark schemes are selected:

\begin{itemize}	
	\item
	\emph{Offloading only scheme}: All tasks of the WSs can only be offloaded to the HAP for edge computing.
	\item
	\emph{Local computing only scheme}: The WSs can only implement their tasks on their local processors.
	\item
	\emph{Passive offloading scheme}: The offloaded tasks can only be transferred to the HAP in the passive offloading phase, and the active offloading phase is not considered.
	\item
	\emph{Active offloading scheme}: The offloaded tasks can only be transferred to the HAP in the active offloading phase, and the passive offloading phase is not considered.
	\item
	\emph{Without STAR-RIS scheme}: No STAR-RIS is deployed for assisting the transmissions between the HAP and WSs.
   \item
	\emph{Reflecting-only RIS scheme}: The reflecting-only RIS is deployed, based on which only the WSs in the reflection space can be aided by the RIS.
\end{itemize}

\begin{figure}[t]
	\centering
	\includegraphics[height=6cm,width=7.5cm]{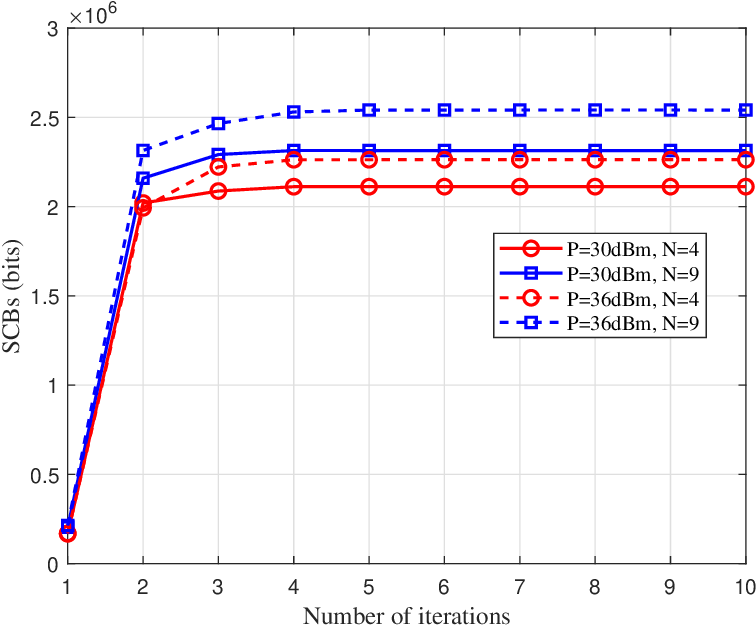}
	\caption{Convergence speed of the proposed Algorithm \ref{alg:pro}.}
	\label{Convergence}
\end{figure}

We first check the convergence speed of Algorithm \ref{alg:pro} in Fig. \ref{Convergence}. As found in Fig. \ref{Convergence}, by implementing Algorithm \ref{alg:pro}, the SCBs first increase with the iterations but quickly converge to a fixed point. This observation verifies that Algorithm \ref{alg:pro} has a superior convergence speed. From Fig. \ref{Convergence}, we also find that the convergence speed is regardless of the HAP's transmit power and the number of antennas, which ensures that Algorithm \ref{alg:pro} can be applicable to different scenarios.

\begin{figure}[t]
	\centering
	\includegraphics[height=6cm,width=7.5cm]{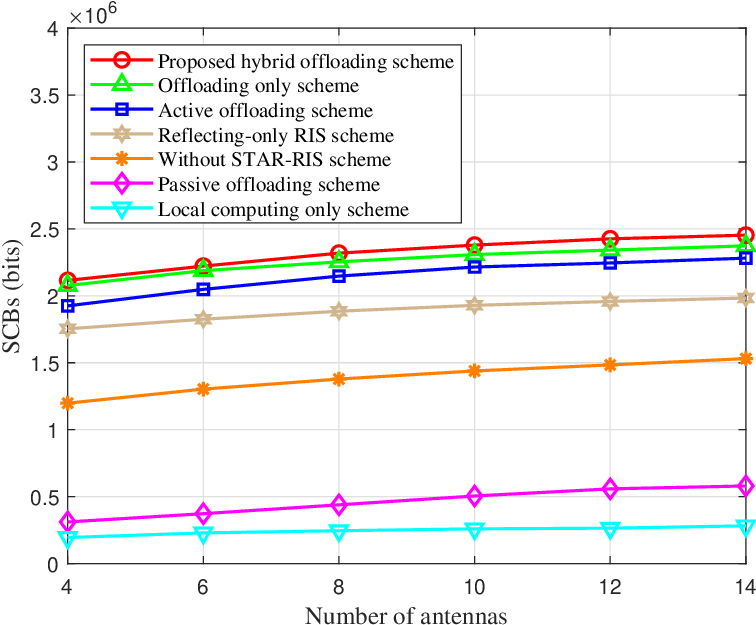}
	\caption{SCBs versus the number of antennas.}
	\label{WD}
\end{figure}

Fig. \ref{WD} illustrates how the SCBs vary with the number of antennas. As expected, the usage of more antennas is an important factor to boost the growth of SCBs. Moreover, the proposed hybrid offloading scheme combing the passive and active offloading achieves the best performance over the benchmark schemes. For example, our proposed scheme demonstrates a performance improvement of 61.1\% compared to the passive offloading scheme and 9.8\% compared to the active offloading scheme. Furthermore, the performance gap between the proposed scheme and the offloading only scheme becomes larger when more antennas are available. It is because the utilization of local computing for executing a portion of tasks is more efficient, especially when the local computing resources are adequate, as opposed to just relying on computing tasks at the MEC server.

\begin{figure}[t]
	\centering
	\includegraphics[height=6cm,width=7.5cm]{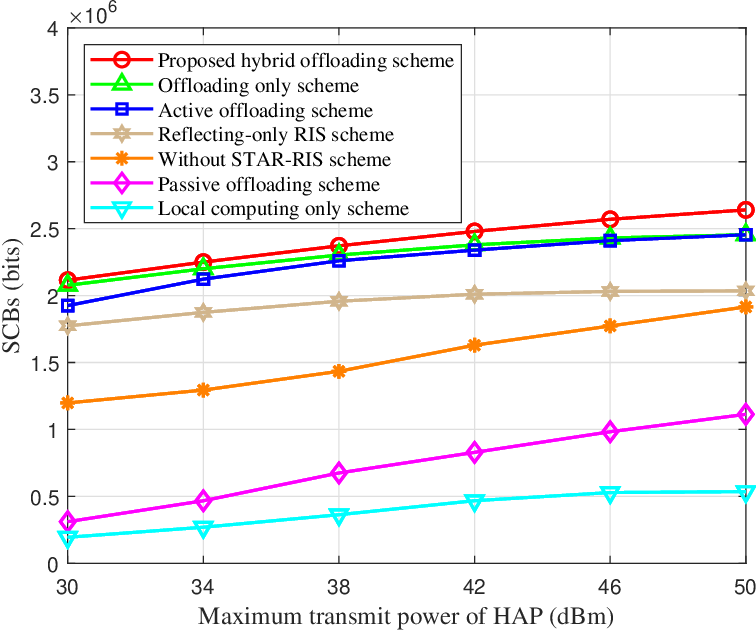}
	\caption{SCBs versus the maximum transmit power of the HAP.}
	\label{power}
\end{figure}

Fig. \ref{power} depicts the SCBs versus the HAP's maximum transmit power (i.e., $P_{\max}$). It is seen that the SCBs is proportional to the maximum transmit power, i.e., transmitting energy signals at the HAP with a large power leads to a performance improvement. It is because more power can be leveraged for supporting the passive offloading and more energy can be harvested for enhancing efficiencies of the active offloading and local computing. The offloading only scheme yields much larger SCBs in comparison to the local computing only scheme. This indicates that the MEC server's computing efficiency is superior as the MEC server typically have a strong computing capability.

\begin{figure}[t]
	\centering
	\includegraphics[height=6cm,width=7.5cm]{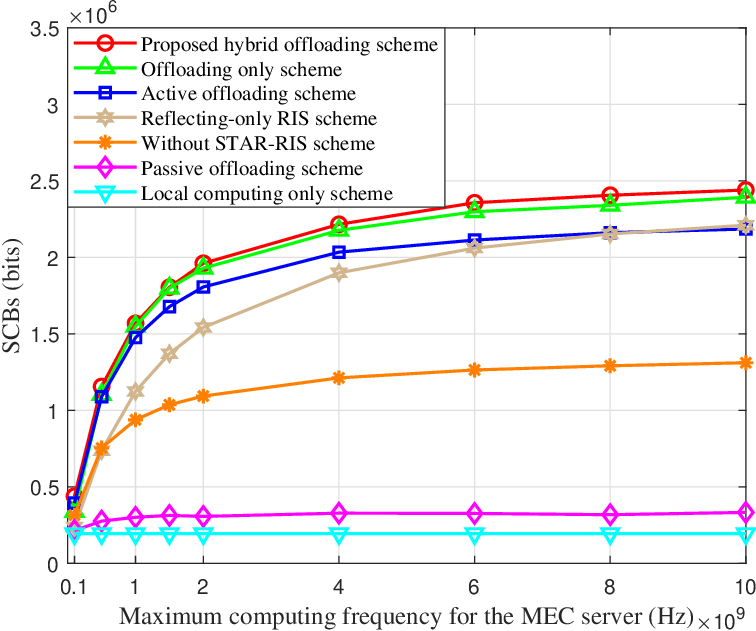}
	\caption{SCBs versus the maximum computing frequency for the MEC server.}
	\label{comp}
\end{figure}

Fig. \ref{comp} presents the SCBs versus the maximum computing frequency (i.e., $f_{\max}$) of the MEC server. It is seen that all schemes involved with edge computing can benefit from increasing the maximum computing frequency of the MEC server. The reason is that with a larger computing frequency, more tasks can be computed per unit time. We also find that when $f_{\max}$ is sufficiently large, the amount of SCBs tends to be stable. It is because even increasing the maximum computing frequency can reduce the edge computing time, the time allocated to passive and active offloading is limited to the block duration. Furthermore, the proposed hybrid offloading scheme demonstrates superior performance compared to the offloading only scheme and the local computing only scheme. This superiority arises from the scheme's ability to enable task computing both locally and remotely.

\begin{figure}[t]
	\centering
	\includegraphics[height=6cm,width=7.5cm]{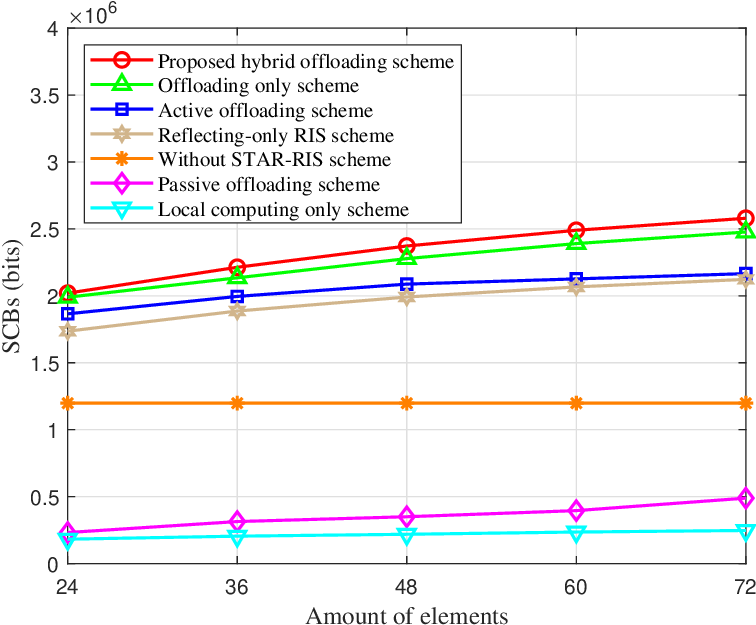}
	\caption{SCBs versus the amount of elements.}
	\label{RE}
\end{figure}

Fig. \ref{RE} depicts the relationship between the amount of elements and the SCBs. It is notable that equipping the STAR-RIS/reflecting-only RIS with more elements shows a positive impact on the increase of the SCBs, which is in line with expectations. This phenomenon is attributed to the fact that more transmission links between the HAP and the WSs can be provided for supporting the passive and active offloading. For instance, in the passive offloading phase, the passive offloading from the WSs to the HAP by reflecting the incident signals from the HAP can be strengthened twice. Similarly, in the active offloading phase, the utilization of the STAR-RIS/reflecting-only RIS enables more tasks to be offloaded to the HAP. However, compared to the reflecting-only RIS scheme, the proposed STAR-RIS scheme can improve SCBs by 21.5\%. This is because unlike the reflecting-only RIS, the STAR-RIS can assist the WSs deployed in both transmission and reflection spaces for computing offloading. Again, the usage of the hybrid offloading scheme ensures superior performance in comparison to the passive offloading scheme and the active offloading scheme. 

\section{Conclusion} \label{V}
In this paper, we have proposed an innovative scheme leveraging the STAR-RIS in WPB-MEC networks, where the STAR-RIS is utilized to ensure comprehensive spatial coverage of all WSs and performance boosting in passive and active offloading phases. By exploring the hybrid offloading characteristics, the HAP can support the FD mode for simultaneous energy transfer and task receiving in the passive offloading phase and focus on receiving tasks by constructing beamforming in the active offloading phase. This dual roles ensure a high efficient utilization of network resources. Furthermore, we have investigated the SCBs maximization problem and proposed a BCD-based algorithm that integrates SCA and SDR techniques to attain a solution with high accuracy. Numerical results have extensively demonstrated the superior performance of our proposed scheme integrating the STAR-RIS and hybrid offloading.
	\bibliography{my}

\end{sloppypar}
\end{document}